\documentclass[tighten,twocolumn]{aastex631}
\usepackage{graphicx}
\usepackage{ulem}
\usepackage{amsmath}
\usepackage{wrapfig}
\usepackage[thinlines]{easytable}
\usepackage{tikz}
\usepackage{hyperref}
\usepackage{xcolor}
\usepackage{ulem}

\begin{document}
\shorttitle{imbh tidal disruption events}
\shortauthors{K{\i}ro\u{g}lu et al.}

\title{Partial Tidal Disruptions of Main-Sequence Stars by Intermediate-Mass Black Holes}

\correspondingauthor{Fulya K{\i}ro\u{g}lu}
\email{fulyakiroglu2024@u.northwestern.edu}

\author[0000-0003-4412-2176]{Fulya K{\i}ro\u{g}lu}
\affiliation{Center for Interdisciplinary Exploration \& Research in Astrophysics (CIERA) and Department of Physics \& Astronomy \\ Northwestern University, Evanston, IL 60208, USA}

\author[0000-0002-7444-7599]{James C.\ Lombardi Jr.}
\affiliation{Department of Physics, Allegheny College, Meadville, Pennsylvania 16335, USA}

\author[0000-0002-4086-3180]{Kyle Kremer}
\affiliation{TAPIR, California Institute of Technology, Pasadena, CA 91125, USA}
\affiliation{The Observatories of the Carnegie Institution for Science, Pasadena, CA 91101, USA}

\author[0000-0002-7330-027X]{Giacomo Fragione}
\affiliation{Center for Interdisciplinary Exploration \& Research in Astrophysics (CIERA) and Department of Physics \& Astronomy \\ Northwestern University, Evanston, IL 60208, USA}

\author[0000-0002-5965-4944]{Shane\ Fogarty}
\affiliation{Department of Physics, Allegheny College, Meadville, Pennsylvania 16335, USA}

\author[0000-0002-7132-418X]{Frederic A. Rasio}
\affiliation{Center for Interdisciplinary Exploration \& Research in Astrophysics (CIERA) and Department of Physics \& Astronomy \\ Northwestern University, Evanston, IL 60208, USA}

\begin{abstract}
We study close encounters of a $1\,M_{\odot}$ middle-age main-sequence star (modeled using MESA) with massive black holes through hydrodynamic simulations, and explore in particular the dependence of the outcomes on the black hole mass. 
We consider here black holes in the intermediate-mass range, $M_{\rm BH}= 100-10^4\,M_{\odot}$. Possible outcomes vary from a small tidal perturbation for weak encounters all the way to partial or full disruption for stronger encounters. We find that stronger encounters lead to increased mass loss at the first pericenter passage, in many cases ejecting the partially disrupted star on an unbound orbit. For encounters that initially produce a bound system, with only partial stripping of the star, the fraction of mass stripped from the star increases with each subsequent pericenter passage and a stellar remnant of finite mass is ultimately ejected in all cases. The critical penetration depth that separates bound and unbound remnants has a dependence on the black hole mass when $M_{\rm BH} \lesssim 10^3\,M_{\odot}$. We also find that the number of successive close passages before ejection decreases as we go from the stellar-mass black hole to the intermediate-mass black hole regime. For instance, after an initial encounter right at the classical tidal disruption limit, a $1\,M_{\odot}$ star undergoes 16 (5) pericenter passages before ejection from a $10\,M_{\odot}$ ($100\,M_{\odot}$) black hole. Observations of periodic flares from these repeated close passages could in principle indicate signatures of a partial tidal disruption event.
\vspace{1cm}
\end{abstract}

\section{Introduction}

Intermediate-mass black holes (IMBHs) in the mass range $10^2-10^5\,M_{\odot}$ \citep[for review, see][]{Greene_2020} are the missing link between stellar-mass black holes ($\lesssim 100\,M_{\odot}$) and super massive black holes (SMBHs; $\gtrsim 10^6\,M_{\odot}$). Despite significant observational and theoretical efforts, the existence of IMBHs is still debated \citep[e.g.,][]{Noyola_2010, Lanzoni_2013, Baumdardt_2017, Kiziltan_2017, Tremou_2018, Pechetti_2022}.

Stellar systems with high central densities $\gtrsim 10^5\, \rm{pc}^{-3}$, such as globular clusters (GCs), have been claimed to yield optimal environments to form an IMBH either via sequential mergers of stellar-mass black holes \citep{Miller_2002, Gurkan_2006, Freitag_2006, O'Leary_2006, Umbreit_2012, Rodriguez_2018, Antonini_2019, Rodriguez_2019, Fragione_2020, Fragione_2020b, Mapelli_2021, Fragione_2022a, Fragione_2022b} or as a result of the collapse of a massive star from stellar collisions and mergers \citep{Ebisuzaki_2001, Portegies_Zwart_2002, Gurkan_2004, Portegies_Zwart_2004, Vanbeveren_2009, Banerjee_2020, Banerjee_2020_b, Kremer_2020, Das_2021, DiCarlo_2021, Gonzalez_2021}.

Accretion signatures provide a viable method to search for IMBHs, and some accreting IMBH candidates have recently been identified in extragalactic young star clusters \citep[][]{Farrell_2012, Webb_2012, Soria_2013, Mezcua_2015}. In particular, multiwavelength follow-up observations of the luminous X-ray flare J1847 in an off-center massive star
cluster suggest that it is a TDE event caused by an IMBH \citep{Lin_2018}. Moreover, in dense star clusters, IMBHs can form binaries that become gravitational wave sources with unique properties, observable with present and upcoming observatories \citep[e.g.,][]{Miller_2002, Mandel_2008, Fragione_2018a, Fragione_2018b, Gonzalez_2022}. These make dense star clusters primary targets to hunt for IMBHs. Yet the existence of an IMBH in dense stellar environments has been controversial. For example, the observable dynamical properties of an IMBH inhabiting a GC, such as velocity dispersion and surface brightness profiles, can be reproduced by effects of rotation \citep[][]{Zocchi_2015} or a stellar-mass black hole subsystem \citep[][]{Zocchi_2017}.

There has been a growing number of studies of TDEs by stellar-mass black holes \citep[][]{Perets_2016, Lopez_2019, Kremer_2019, Kremer_2021, Kremer_2022, Wang_2021, Ryu_2022} and especially SMBHs \citep[for review, see][]{Stone_2019}. Some studies have also investigated the TDE of a white dwarf \citep[][]{Lu_2008, Rosswog_2009, Krolik_2011, Haas_2012, MacLeod_2016} or a main-sequence (MS) star \citep[][]{Ramirez_Ruiz_2009, Chen_2018} by an IMBH and discussed their observational signature. Particularly, very energetic TDEs by IMBHs have been proposed as one potential way to result in jet formation and produce a long gamma-ray burst \citep[][]{Lu_2008}.

The strength of a tidal encounter is parameterized by the so-called penetration factor $\beta \equiv r_T/r_p$, where $r_p$ is the pericenter distance (distance of closest approach) and $r_T = (M_{\rm BH}/M_{\star})^{1/3} R_{\star}$ is the classical tidal disruption radius \citep[][]{Press_1977}, where $M_{\star}$ and $R_{\star}$ are the stellar mass and radius, respectively. The tidal disruption radius defines the distance over which the black hole’s tidal force exceeds the star’s self-gravity \textit{at the stellar surface}. This means that encounters with $r_p \leq r_T$ do not necessarily yield full disruption events, especially for stars with a high central density. This naturally raises the question of how the “physical tidal radius” (i.e., maximum pericenter distance for a full disruption) changes with the internal stellar structure of a realistic star. In the context of these searches, \cite{Guillochon_2013} found the critical pericenter distance for the complete disruption of a $1\,M_{\odot}$ mass star modeled with a polytrope $\gamma=4/3(5/3)$  to be $r_T/r_p=2(0.6)$.
The partial and complete disruption of stars has also been studied \citep[][]{Golightly_2019, Law-Smith_2019, Law_Smith_2020, Ryu_2020a, Ryu_2020b, Ryu_2020c} by using realistic stellar structures with the MESA \citep{Paxton_2011} stellar evolution code. The studies of these TDEs, however, are mainly focused on the SMBH regime with a black hole mass $M_{\rm BH} \gtrsim 10^5\,M_{\odot}$.

In the case of a partial disruption event, the stellar remnant may be either unbound (i.e., orbital energy becomes positive) or bound to the black hole depending on the details of the mass loss. Previous studies have shown that asymmetric mass ejecta results in an impulsive ``kick'' in the context of planets \citep[e.g.,][]{Faber_2005, Guillochon_2011, Liu_2013}, MS stars \citep{Manukian_2013, Kremer_2022}, white dwarfs \citep{Cheng_2013}, and neutron stars \citep{Rosswog_2000, Kyutoku_2013}. In particular, \cite{Manukian_2013, Gafton_2015} studied kicks resulting from the TDEs of stars by SMBHs in galactic nuclei. Their smoothed-particle hydrodynamics (SPH) simulations show that the kick velocity the star receives (which can be as large as the star’s own escape velocity) is \textit{independent} of the mass ratio in the range $10^3$--$10^6$ but depends on the amount of mass lost at the pericenter.

Building off earlier work that explored specific regions of the IBMH TDE parameter space, in this paper, we present the results of the first systematic study on the close encounters of a MS star by IMBHs through a large set of hydrodynamic simulations. We follow up on our recent study, \cite{Kremer_2022}, in which we investigate the TDEs of stars by stellar-mass black holes using polytropic and Eddington standard stellar models. In this work, we perform SPH simulations using a wider range of black hole masses but mainly focus on the IMBH regime, and update our stellar models using the stellar evolution code MESA \citep[][]{Paxton_2011}.  Similar to previous works, our study of the disruption of a MESA stellar model by an IMBH reveals partially-disrupted stars for a range in $r_p$ values, which depends on the stellar density profile. Different from previous studies, we also explore how the boundedness of partially-disrupted cores, kick velocities, and number of pericenter passages before ejection vary with black hole mass for a given star.
 
 This paper is organized as follows. In Section \ref{methods}, we describe the computational method and the stellar profile obtained with MESA. In Section \ref{results}, we present the outcomes of hydrodynamic calculations and discuss how varying black hole mass and pericenter distance affect the outcomes. We also determine the properties of the accretion disks formed and discuss possible electromagnetic signatures. Finally, we conclude and discuss our results in Section \ref{discussion}.
 
\section{Methods}\label{methods}
\begin{figure*}
    \centering
    \includegraphics[width=0.82\linewidth]{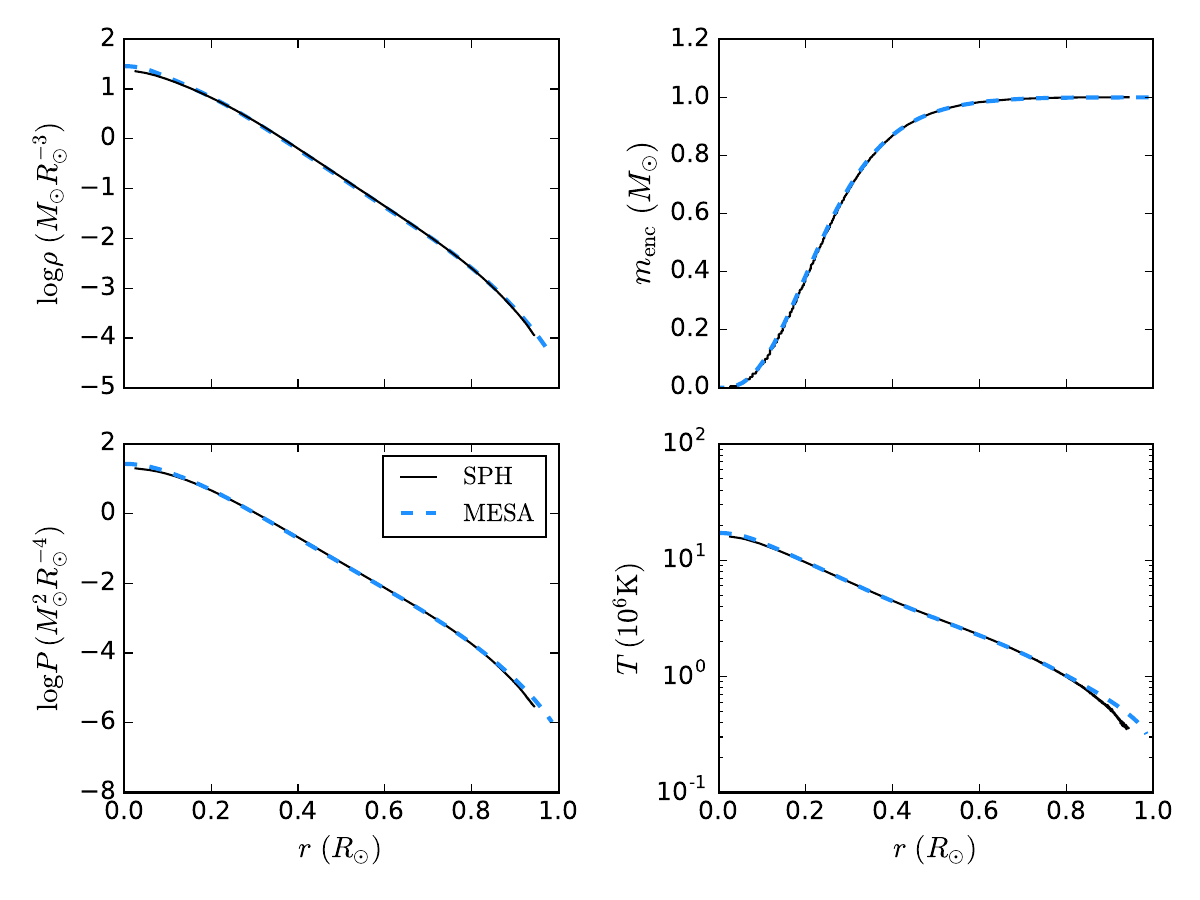}
    \caption{\footnotesize Comparison of the 1D MESA model (dashed lines) with the SPH model of the star at the end of relaxation (solid lines). The density, mass, pressure, and temperature profiles of the SPH model agree well with the MESA model. }
    \label{fig:MESA_vs_SPH}
\end{figure*}
We have computed a series of close encounters between a black hole and a MS star of mass $1M_{\odot}$ using the SPH code \texttt{StarSmasher}\footnote{\texttt{StarSmasher} is available at \href{https://jalombar.github.io/starsmasher/}{https://jalombar.github.io/starsmasher/}} \citep{RasioThesis, Gaburov_2010b, Gaburov_2018}. SPH is a Lagrangian particle method in which the star is represented as a collection of fluid particles, each carrying a mass $m_i$, position $r_i$, velocity $v_i$, and specific internal energy $u_i$. Each particle also has an associated smoothing length $h_i$ that determines the local spatial resolution. The code implements a Wendland C4 smoothing kernel, as described in \cite{Wendland_1995}, as well as an artificial viscosity prescription coupled with a Balsara Switch \citep[][]{Balsara_1995} to prevent unphysical inter-particle penetration. Gravitational forces and energies are computed by performing a direct summation on NVIDIA graphics cards as described in \cite{Gaburov_2010b}, which has been shown to be more accurate than a tree based method at the cost of speed \citep{Gaburov_2010a}. We direct the reader to \cite{Kremer_2022} for a detailed description of \texttt{StarSmasher}'s implementation of all these processes, including prescriptions for smoothing length determination and time-stepping.

For all calculations, we adopt a stellar mass of $1\,M_{\odot}$ while black hole masses range from $10\,M_{\odot}$ to $10^4\,M_{\odot}$. 
In each SPH run, the star is initially placed on a parabolic incoming trajectory ($v_{\infty}=0~\rm{km}/{s}$) toward the black hole with an initial separation $10~r_T$, where tidal effects are negligible. Given this, the first pericenter passage of the star occurs $\approx 0.3$ days after the simulation starts. We treat the black hole as a (softened) point mass interacting only gravitationally with the SPH particles and set its constant gravitational softening length to be the root mean square of the smoothing lengths of the SPH particles in the parent star.  Except in the $r_p/r_T >1.1$ cases, all of the simulations are run until the most bound material returns to the pericenter after the last passage, hits the original tidal stream, and circularizes (thus forming a thick rotating torus or ``disk") on a timescale comparable to the period of that material: 
\begin{align}  
    \label{eq:t_fb}
    t_{\rm orb} &= 2\pi \sqrt{\frac{\max(r_p^3,r_T^3)}{G M_{\rm BH}}} \nonumber \\ &\approx 10 \Bigg( \frac{M_{\rm BH}}{100\,M_{\odot}} \Bigg)^{-1/2} \Bigg( \frac{r_p}{4.6\,R_{\odot}} \Bigg)^{3/2} \, \rm{hr} 
\end{align}
where $r_p=4.6\,R_{\odot}$ is the tidal disruption radius of a $1\,M_{\odot}$ star for a $100\,M_{\odot}$ black hole.
We note that the orbital period approximation $t\propto r_p^{3/2}$ is appropriate when $M_{\rm BH}\lesssim 10^3\,M_{\odot}$ and differs from the characteristic fallback timescale $t\propto r_p^3$ \citep[e.g.,][]{Lacy1982,Rees_1988,EvansKochanek1989} which is appropriate for $R_{\odot} \ll r_p$ as in the supermassive black hole TDE regime.

\subsection{Stellar models}

We prepare an SPH model of the star in isolation before initiating a collision. Our study employs a realistic star model created with MESA \citep{Paxton_2011}. We evolve the 1D MESA star with initial mass $M_{\star}=1\,M_{\odot}$, helium abundance $Y=0.24$ and metallicity $Z=0.001$ until it reaches its half MS lifetime $\sim~2.5\times 10^9$~yr. The ratio of the core $\rho_c$ and the average density is about 150. Next, in order to convert the detailed stellar evolution model into SPH initial conditions, we interpolate particle density, pressure, and mean molecular weight at a particle distance $r_i$ from the 1D model as described in e.g., \citet{Sills_2001, Hwang_2015}. We generate a 3D SPH model of the 1D model using $N=10^5$ unequal-mass particles, with an average number of neighbors of $N_N\sim150$. Because we are primarily interested in unbound ejecta mass after the pericenter passage, we focus on the hydrodynamics of the initial interactions. Hence, we provide higher resolution near the stellar surface to better resolve the onset of tidal disruption. The SPH particles are initially placed on a hexagonal close-packed lattice, which is stable to perturbations \citep{Lombardi_1999}. After initialization of the SPH particles, we allow the SPH fluid to settle into hydrostatic equilibrium by adding an artificial drag force to the hydrodynamical accelerations to dampen the oscillations.  The SPH particles are advanced using the variational equations of motion \citep{Monaghan_2002, Springel_2002, Lombardi_2006}. Radiative cooling and heating are neglected in this study. 

In Figure \ref{fig:MESA_vs_SPH}, we compare the pressure, temperature, density, and enclosed mass profiles of the relaxed SPH and MESA model as a function of radius. We note that the radius of the relaxed star mapped into three dimensions is smaller than that of the MESA star. This is because we define the stellar radius as the position of the outermost SPH particle rather than the photosphere radius of the MESA star. In this case, the center of the outermost particle is at $\approx 0.9\,R_{\odot}$, with a smoothing kernel that extends out an additional distance $2\,h_{\rm out}$ from the center of the particle, where $h_{\rm out}\approx 0.07\,R_\odot$ is the smoothing length of the outermost particle. See Section 2.2 and Table 1 of \cite{Nandez_2014} for a discussion and alternative ways of measuring the radius of an SPH star model. Apart from this discrepancy, the relaxed model agrees reasonably well with the desired MESA profile.  
We calculate the pressure at each particle radius using the radiation pressure and ideal gas contributions as
\begin{equation}
    P_i = \frac{\rho_i k T_i}{\mu_i}+\frac 1 3 a T_i^4,
\end{equation}
where $T_i$ is temperature, $\mu_i$ is mean molecular mass, $a$ is the radiation constant and $k$ is the Boltzmann constant. We find the temperature for each particle by solving the fourth order equation
\begin{equation}
    u_i = \frac 3 2 \frac{k T_i}{\mu_i} + \frac{aT_i^4}{\rho_i},
\end{equation}
which comes from $u = u_{\rm gas}+ u_{\rm rad}$, with $u_{\rm gas}$ proportional to $T$ and $u_{\rm rad}$ proportional to $T^4$.

Because we use a constant number density in the parent model, the particle masses $m_i$ are assigned to follow the desired density profile, $m_i \propto \rho(r_i)$, and normalized to give the desired stellar mass.
As compared to a model with equal-mass particles, this parent star has more particles in its outer regions, which increases the resolution near the surface. Models with equal-mass particles, in which particles are more closely packed in the center of the star than near the surface, are better to use for nearly head-on interactions ($r_p \sim 0$), which lead ultimately to mergers. In the more general case of off-axis interactions, however, the resulting disk formed around the black hole primarily consists of particles from the outer portions of the parent star, favoring placing more particles there. 

\subsection{Orbital regularization} \label{orbital_jumps}

Some of the grazing encounters between a star and a black hole result in a bound orbit around the black hole with a high eccentricity  $0.96 < e < 1$. In the case of $M_{\rm BH}>100\,M_{\odot}$ and $r_p=r_T$, the eccentricity is larger than 0.999 and the semi-major axis $a$ is more than $10^4\,R_{\odot}$, corresponding to an orbital period of at least $\sim 10$ years or $\sim 10^5$ dynamical timescales. Hence, for computational efficiency in such situations, we use the analytic solution to the Kepler two-body problem to advance the orbit of the bound star to the point with the same separation from the black hole but now infalling toward it \citep{Antonini_2011, Godet_2014}.

Since our primary focus is determining how much mass falls to the black hole with each passage, we do not wait for the disk to reach a steady state before performing such an orbital jump. Instead, we run (or rerun) all our simulations resulting in a bound star with an orbital jump once the star has receded to a separation exceeding  $r_{\rm jump} \equiv 1.7\,(M_{\rm BH}/M_{\star})^{1/3}\,{\rm \max}(r_{\rm p},r_{\rm T})$, corresponding to a sufficiently late time that the orbital parameters are well-determined and the star is close to hydrostatic equilibrium.  
For $r_{\rm p} \gtrsim r_{\rm T}$, this separation is reached
when the debris stream is just starting to self-intersect. Because we are not considering shocks in this subset of runs, we turn off artificial viscosity so that oscillations in the remnant star can be followed as accurately as possible. In many cases, the viscous accretion time of the disk formed during each passage is less than or comparable to the orbital period for the partially disrupted remnant to return. We, therefore, make the approximation that all of the debris bound to the black hole has its mass and momentum added to the black hole when the orbit jumps ahead. This approximation may not always be ideal, as the orbital time can be comparable to or even smaller than the accretion timescale in the stellar-mass black hole cases \citep{Kremer_2022}. However, because $M_{\rm BH} \gg M_*$, the black hole mass does not change appreciably even in this scenario of unadulterated accretion and so our results are not sensitive to this approximation. Additionally, all the material stripped from the star is thrown away to infinity during the advancement of the orbit.

\section{Results}\label{results} 

\begin{figure}[!t]
    \centering
    \includegraphics[width=\linewidth]{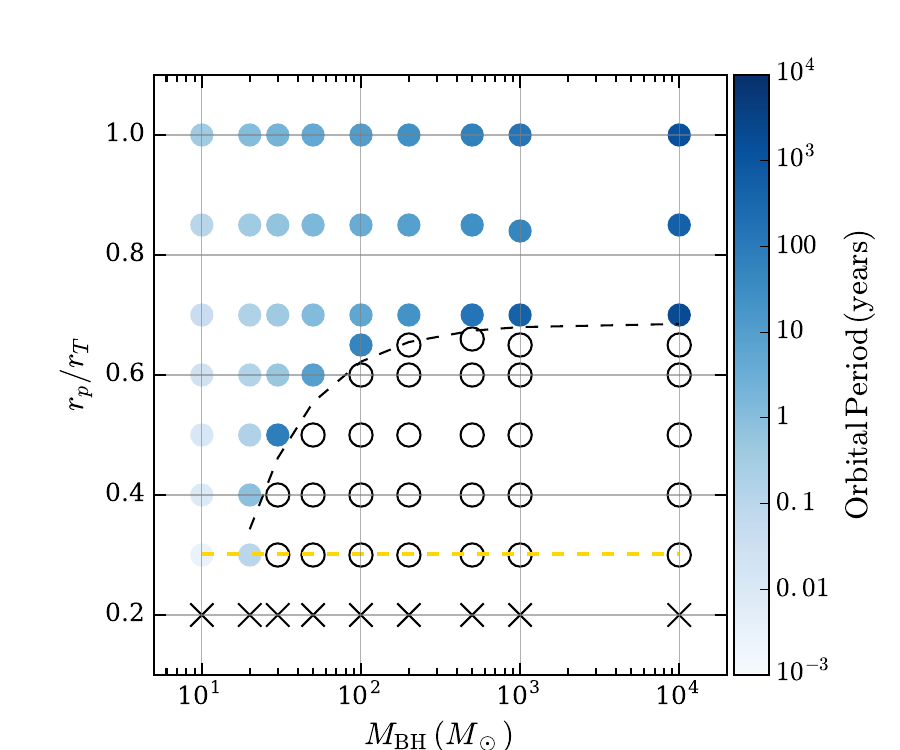}
    \caption{\footnotesize
   Outcomes of the simulations performed with various black hole masses and pericenter distances after the first pericenter passage. The complete tidal disruption results are indicated by X's. The empty and blue-filled circles denote the partial tidal disruption events leaving behind an unbound and bound remnant, respectively. Blue shading represents the binary orbital period when the star is bound to the black hole. The dashed black line shows the boundary below which the star receives an impulsive kick and becomes unbound. The yellow dashed line indicates the boundary between full and partial disruption events obtained with a semi-analytical approach in Section \ref{boundary_sec}. All models adopt $1M_{\odot}$ star at the MS middle age. 
    \label{fig:simulation_results}}
\end{figure}

In this section, we report the results of hydrodynamic calculations of parabolic encounters between a $\,1M_{\odot}$ MS star and a black hole. In Table \ref{first_passage}, we list all models with the initial conditions in Columns 1-2 and outcomes after the \textit{first} pericenter passage in the remaining columns. We vary the pericenter distance between $0.2-1.75\,r_T$, covering a broad range of outcomes. In our recent work \citep{Kremer_2022}, we showed that these types of events can lead to a fully disrupted star or partially disrupted bound/unbound remnant depending on the distance of the closest approach. Here, in Figure \ref{fig:simulation_results}, we similarly demonstrate the three outcomes after the first pericenter passage. For the strongest interactions with $r_p/r_T < 0.3$ independent of black hole mass, the star is fully disrupted. For $r_p/r_T > 0.3$, the star is partially disrupted with its dense core remaining intact while the outer envelope is stripped off. Depending on the orbital energy of the remnant star, it can then either form a (bound) binary with the black hole or remain unbound receiving an impulsive kick from an asymmetric mass loss at the pericenter.

In Column 3 of Table \ref{first_passage}, we list the outcomes for a fully disrupted (F), unbound (U), and bound (B) star.
In Columns 4–6, we show, respectively, the total mass bound to the black hole $M_{\rm BH,bound}$, the mass of the star (which is zero in the case of a full disruption), and the total mass of material that has been unbound entirely from the system $M_{\rm ej}$. The U and B outcomes are reported after the first pericenter passage when the separation between the star and black hole exceeds the jump radius $r_{\rm jump}$. Because there is no obvious way to define a separation once the star is destroyed, we report the F cases at a fallback time $t_{\rm fb}$  after the first pericenter passage, which is given by
\begin{equation}  
t_{\rm fb} = \frac{2 \pi}{(G M_{\rm BH})^{1/2}(2 R_{\star})^{3/2}}\times \rm{max}[r_p^3,r_T^3],\label{t_fb}
\end{equation}
which increases for $r_p > r_T$. In Table \ref{last_passage}, we list the results of all the simulations at $t_{\rm fb}$ after the \textit{last} pericenter passage.

In the following subsection, we identify the boundary between full and partial disruptions, and partial disruptions resulting in an unbound and bound star.

\subsection{The boundary between total and partial disruption} \label{boundary_sec}

\begin{figure}
    \centering
    \includegraphics[width=1\linewidth]{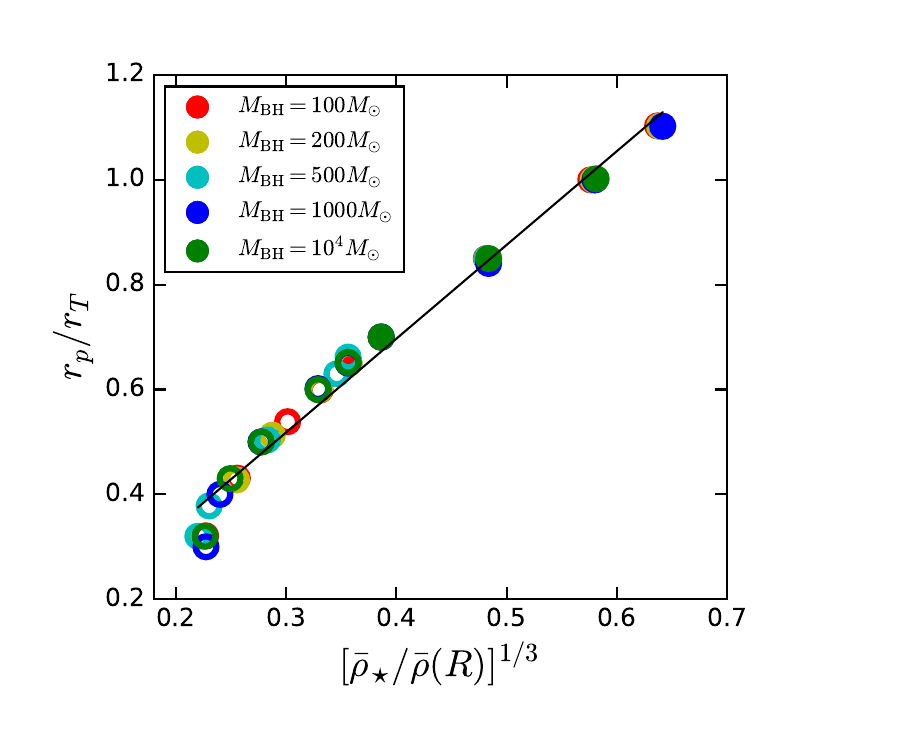}
    \caption{\footnotesize  The linear relation between the density ratio and penetration factor for the partially disrupted star models, which gives $\zeta \simeq 5.6$. For each remnant mass $M(R)$, we find the radius $R$ from the MESA profile. Different colors denote the black hole mass. The filled circles denote models where the stellar remnant is tidally captured by the black hole, forming a bound binary. Open circles denote models where the partially disrupted star remains unbound after the first pericenter passage.}
    \label{fig:density}
\end{figure}

Equating the self-gravity of the star to a factor ($\zeta$) times the tidal force applied by the black hole at the radius $R$ enclosing the remnant mass $M(R)$, one can obtain a simple expression for the radius beyond which mass in the star is stripped \citep[][]{Law-Smith_2019,Ryu_2020b}:
\begin{equation}
   \beta^{-1}= \frac{r_p}{r_T}= \zeta^{1/3} \left(\frac{\bar{\rho}_{\star}}{\bar{\rho}(R)}\right)^{1/3} \label{eq:boundary}
\end{equation}
where $\bar{\rho}_{\star}=3M_{\star}/(4\pi R_{\star}^3)$ and $\bar{\rho} = 3M(R)/(4\pi R^3) $. For each remnant mass $M(R)$, we find the radius $R$ from the MESA profiles and get the coefficient $\zeta \simeq 5.6$ by fitting the linear relation between the density ratio and $r_p/r_T$ as shown in Figure \ref{fig:density}. Similar to the work of \cite{Ryu_2020b} and \cite{Law_Smith_2020}, we fit the critical penetration factor at which the star is fully disrupted in the limit $R \rightarrow 0$ at which $\bar{\rho} = \rho_c$. For $\bar{\rho}_{\star}/\rho_c \simeq 0.0067 $ and  $\zeta \simeq 5.6$, we find the critical $r_p/r_T$ to be $\beta_c^{-1} \simeq 0.3$. In Figure \ref{fig:simulation_results}, we show the boundary between full and partial disruption events for each model with a dashed yellow line which agrees well with the SPH results. Our results are in agreement with the hydrodynamic simulations \cite{Law_Smith_2020} performed with FLASH based on a MESA model for the solar-mass MS star at middle age, finding $\beta_c^{-1} \simeq 0.37$. In the case of $M_{\rm BH}\sim 10^6\,M_{\odot}$, \cite{Ryu_2020b} show that full tidal disruptions can occur for larger pericenter distances due to the relativistic effects and find the critical penetration factor to be $\beta_c^{-1}\simeq 0.47$.

\subsection{Kick velocities}

After partial disruption, the remnant core either returns to the black hole to be disrupted again or escapes from the system. This bound/unbound dichotomy in the surviving stellar remnant's fate after the first pericenter passage depends on the remnant's orbital energy loss or gain. As described in \cite{Kremer_2022}, the total orbital energy of the star after the first pericenter passage can be written as
\begin{equation}
    E_{\rm orb} = E_{\rm tides}+E_{\rm bind}+E_{\rm kick},
        \label{eq:eorb}
\end{equation}
where $E_{\rm tides}$ is the injection of orbital energy into oscillation modes of the star through tides, $E_{\rm bind}$ is the binding energy of the disrupted star and ejected material, and $E_{\rm kick}$ is the kinetic energy imparted to the stellar core from the mass loss. These quantities are given in Equations (8), (10), and (11) in \cite{Kremer_2022}.  Although the tidal energy equation is designed to treat weak interactions, it is applied here to offer a simple estimate of tidal effects, even in cases where the significant mass loss occurs.
In partial disruption events, the material stripped from the star is ejected in two tidal tails. While the tail on the side of the star facing the black hole is bound to the black hole, the other is ejected from the system on a hyperbolic trajectory. An asymmetry between these tidal tails can result in a positive change in the orbital energy of the star if $E_{\rm kick}>|E_{\rm tides}+E_{\rm bind}|$. In that case, the net effect is an unbound star with a high-velocity kick. This outcome is qualitatively most likely to occur when the total amount of ejected material from the star is comparable to or larger than the mass of the surviving remnant. On the other hand, for larger penetration depths, the partially disrupted may be captured by the black hole if the removal of the orbital energy to excite oscillations in the star exceeds the injection of energy into the star’s orbit from the asymmetric mass loss. In that case, the change in the orbital energy is negative (any kick imparted to the core is small) and the star is bound to the black hole.

Here we explore the critical distance $r_p/r_T$ at which the star receives a repulsive kick for a given black hole mass by making a power-law fit to our simulation results
\begin{equation}
    \beta_{\rm kick}^{-1}  = \xi \left[1-\left(\frac{M_{\rm BH}}{M_{\rm BH,c}}\right)^{-\alpha}\right] \, \mathrm{for} \, \, M_{\rm BH} > 20\,M_{\odot}.
    \label{eq:beta2}
\end{equation}

 We get the fit parameters as $\xi \approx 0.7$, $\alpha \approx 1.0$ and $M_{\rm BH,c}\approx 10\,M_{\odot}$ using the simulations results near the boundary between unbound and bound partial disruptions. The resulting curve is shown with a dashed black line in Figure \ref{fig:simulation_results}. We never get a partial disruption event resulting in an unbound remnant star for $M_{\rm BH} \leq 20\,M_{\odot}$, where we transition directly from capture to full disruption and thus Equation \ref{eq:beta2} does not apply. This result is in agreement with our previous work \citep{Kremer_2022}, where the $1\,M_{\odot}$ star with an Eddington standard model ($n=3$) is bound to the $10\,M_{\odot}$ black hole after the first pericenter passage in any of the models regardless of the distance of closest approach. 
 
As can be seen in Figure \ref{fig:simulation_results}, the critical penetration depth that separates bound vs unbound remnants has a dependence on the black hole mass when $M_{\rm BH} < 10^3\,M_{\odot}$. This dependence can be obtained quantitatively by inserting $r_p=r_T/\beta$ into the total orbital energy (Equation~\ref{eq:eorb}), which gives $E_{\rm orb}\propto 1/M_{\rm BH}^{2/3}$. At the boundary between bound/unbound star, solving for $E_{\rm orb} = 0$ gives a critical penetration depth $\beta$ that depends on $M_{\rm BH}$ for black hole masses less than $\sim 10^3\,M_{\odot}$ but is independent of $M_{\rm BH}$ for higher black hole masses. Indeed, for $M_{\rm BH}\gtrsim 1000\,M_{\odot}$, the critical penetration factor settles to $\beta_{\rm kick}^{-1} \approx \xi \approx 0.7$. From this, we can compare our results from previous work which considered more massive black holes ($M_{\rm BH}\sim 10^6\,M_{\odot}$). \cite{Ryu_2020c} also studied partial disruption of main-sequence stars generated by MESA. Comparing their unbound remnants starting to happen around $r_p/r_T=0.7$ (see their Figure 4) with our results, we find good consistency. 
 
 We list the kick velocities imparted to the star and black hole on the last two columns of Table \ref{first_passage} and \ref{last_passage}. When the star happens to be bound to the black hole, the center of mass of the black hole + star binary receives a momentum kick (tens of kilometers per second at most) in the direction opposite the net momentum of the ejecta. In this case, the kick velocities $v_{\rm kick,\star}$ and $v_{\rm kick,BH}$ listed on Table \ref{first_passage} correspond to the velocity of the center of mass of the binary and therefore have the same value. When $M_{\star}\ll M_{\rm BH}$, the center of mass velocity is negligible as expected. The kick similarly can be imparted to an unbound stellar remnant due to an asymmetric ejection of the material from the star itself due to the action of the black hole. In this particular case, the kick velocities are potentially high enough to eject the star from a GC where the typical escape speed is $\approx 50~\rm{km/s}$. 
 
\begin{figure*}
    \centering
    \includegraphics[width=1\linewidth]{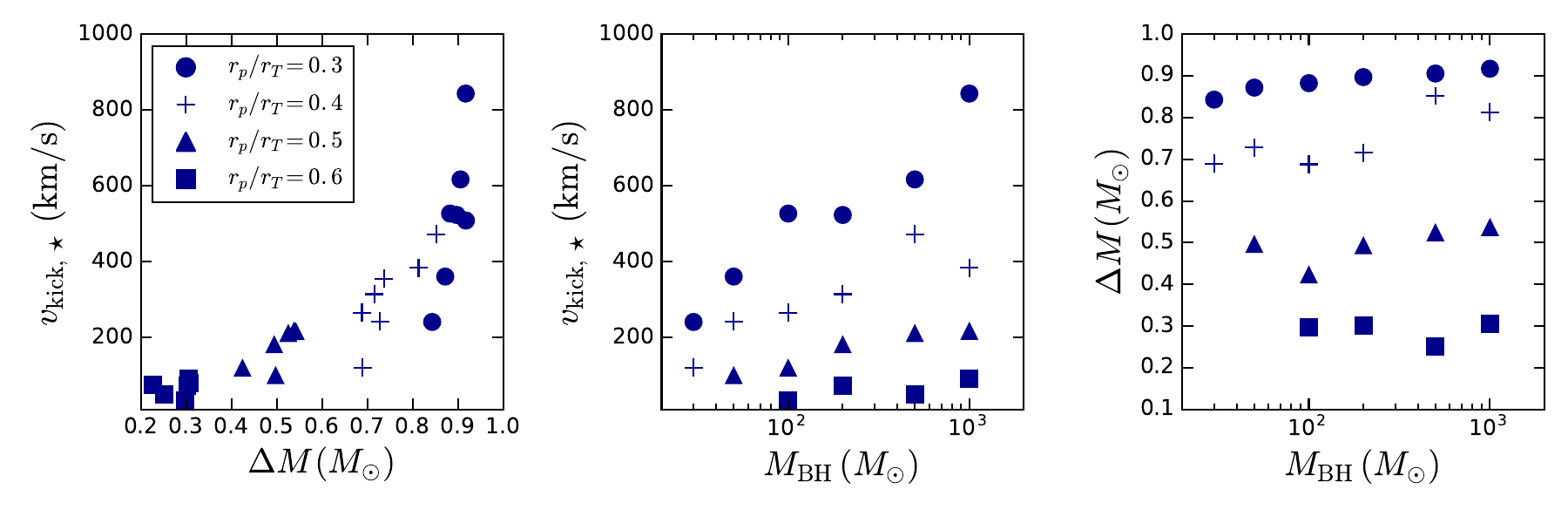}
    \caption{\footnotesize In the left panel, we show the velocity kick the star receives after the first pericenter passage as a function of mass stripped from the star ($\Delta M$). In the middle panel, we show velocity versus $M_{\rm BH}$. In the right panel, we show mass stripped versus $M_{\rm BH}$. Various pericenter distances are indicated by different symbols. Stronger encounters lead to more mass loss and thus larger kicks. Kick velocities also increase with black hole mass even though the mass stripped from the star remains roughly the same at a given $r_p/r_T$.}
    \label{fig:kick_velocities}
\end{figure*}

On the first two panels of Figure \ref{fig:kick_velocities}, we show the kick velocity the unbound star receives after the first pericenter passage as a function of  total mass stripped from the star and black hole mass, respectively. We also consider various pericenter distances in the range $0.3<r_p/r_T<0.6$, which are indicated by different symbols. We see that kick velocities are in general higher for more penetrative encounters, i.e., smaller $r_p/r_T$, which leads to higher mass loss. Additionally, we find that the larger black hole mass case does eject mass at a much larger speed and thus the star receives a larger kick. For instance, at the pericenter distance $r_p=0.3\,r_T$, the kinetic energy, $T$, of the ejecta is about a factor of 6 larger in the $M_{\rm BH}=1000\,M_{\odot}$ case than in the $M_{\rm BH}=50\,M_{\odot}$ case. Neglecting changes in the distribution of the direction of motion for the ejecta from case to case, we expect the momentum of the ejecta to satisfy $T\sim p^2/M_{\rm ej}$. In other words, we expect the momentum carried away to scale as $\sqrt{M_{\rm ej} T}$. Thus, the $M_{\rm BH}=1000\,M_{\odot}$ case with the ejecta mass $M_{\rm ej}\approx 0.5\,M_{\odot}$ should give a remnant speed that is about twice the remnant speed in $M_{\rm BH}=50\,M_{\odot}$ case with $M_{\rm ej}\approx 0.6\,M_{\odot}$, consistent with the data from Table \ref{first_passage}. On the third panel of Figure \ref{fig:kick_velocities}, we show that the mass stripped from the star remains roughly the same by increasing $M_{\rm BH }$ at a given $r_p/r_T$. However, a larger kick is imparted to the star from a more massive black hole due to a larger ejection speed of the stripped material.

\begin{figure*}[t]
   \centering
  \includegraphics[width=1\linewidth]{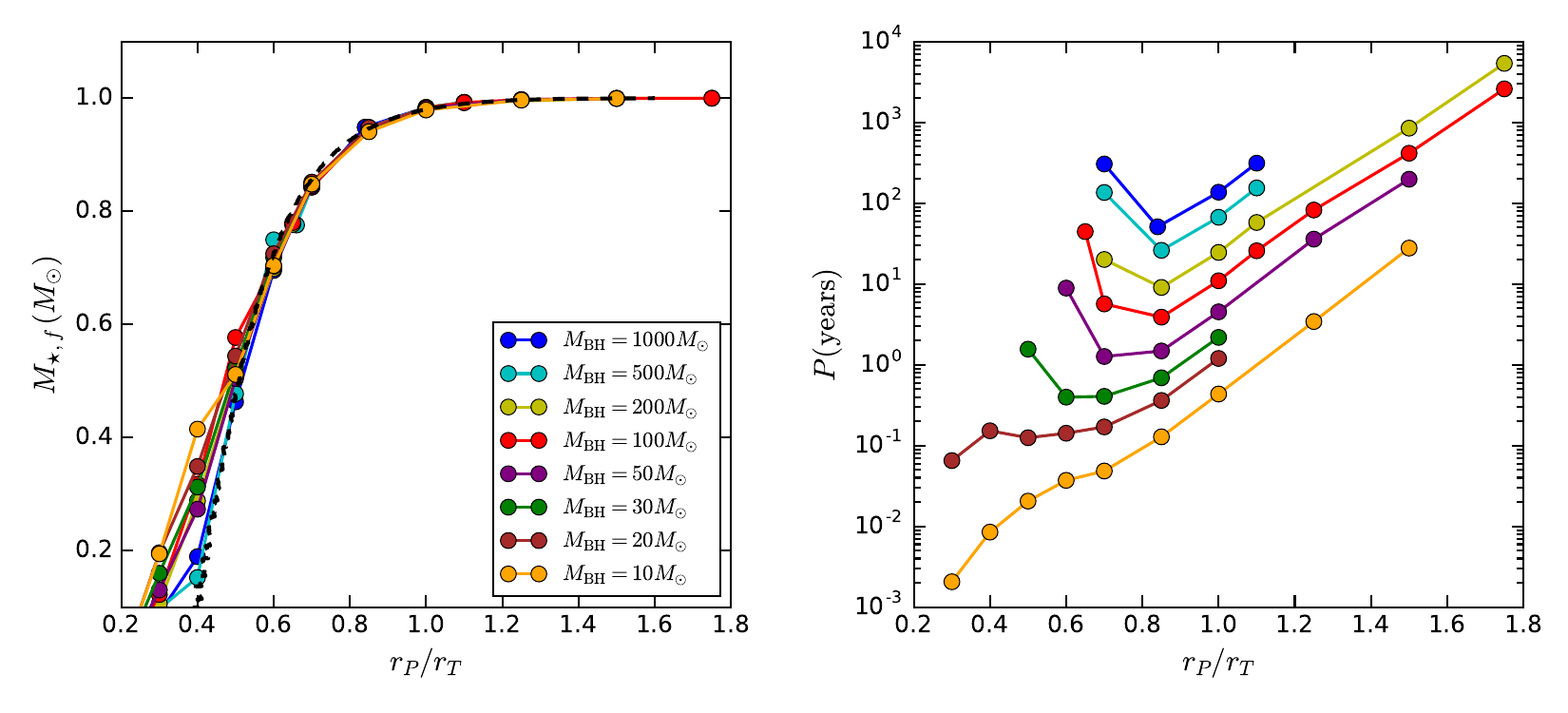}
    \caption{\footnotesize Stellar remnant mass (left panel) and orbital period (right panel) after the first pericenter passage for encounters between $1\,M_{\odot}$ star and the $M_{\rm BH}$ at different pericenter distances. In both panels, various black hole masses are indicated by different colors. Taking $\zeta=5.6$ in Eq \ref{eq:boundary}, we find the required pericenter distance for a given enclosed mass to survive
the encounter at different radii using the MESA
profile of the star. We show the resulting curve with the
black dashed line on the left panel. Even though it underestimates $M_{\star,f}$ at smaller $r_p/r_T$, showing that it is harder to remove mass than our model predicts, it overall agrees well with the SPH results. This shows that the method of \cite{Law-Smith_2019,Ryu_2020b} can be used to predict the remnant mass for a given penetration depth from the density profile of a star without having to run hydrodynamic simulations. Right after the boundary between unbound and bound outcomes, the orbital period is high and then reaches a minimum. As the pericenter distance increases further, less and less mass is stripped and, as a result, the orbital period begins to increase as expected in the classical tidal capture regime.}
    \label{fig:period}
\end{figure*}

 On the left panel of Figure \ref{fig:period}, we show the survived remnant mass $M_{\star,f}$ as a function of $r_p/r_T$ with various $M_{\rm BH}$ after the first passage.  While stronger encounters in more mass loss, the mass stripped from the star starts to be negligible at $r_p/r_T > 1$. Furthermore, we end up getting roughly the same remnant mass at a given $r_p/r_T$ independent of $M_{\rm BH}$. This is because the radius of the star beyond which mass is stripped due to tidal effects depends on $r_p/r_T$ and the density profile of the star.  
 On the right panel of Figure \ref{fig:period}, orbital periods of the bound star after the first pericenter passage  versus $r_p/r_T$ are plotted for different $M_{\rm BH}$. Interestingly, these curves exhibit a minimum for $M_{\rm BH} > 20 M_\odot$ due to the competing effects of the orbital kick due to mass loss and the orbital tightening due to tidal energy transport.
 The larger mass loss in the cases at smaller $r_p$ gives a boost to the semi-major axis that overcomes the decrease in the semi-major axis due to tidal effects. Indeed, at the boundary between bound and unbound results, where the kick balances tides and binding energy changes, the orbital period is infinite. As the pericenter distance increases a little, the system is barely bound so the period is large but finite. At larger $r_p$ still, we eventually get to the minimum of the curve and then enter the classical tidal capture regime, where the period keeps increasing with $r_p$. Since we do not have any unbound result for $M_{\rm BH}=10, 20\,M_{\odot}$, their periods do not have a minimum  in the figure.

\subsection{Multiple passages}

When the star is bound to the black hole after the first pericenter passage, multiple passages may ultimately lead to ejection or inspiral of the core depending on the mass ratio and pericenter distance. Modeling these additional passages, especially in the case of a very long period (e.g., months or more; see Figure \ref{fig:multiple_peri}), is computationally expensive. Hence, as explained in Section \ref{orbital_jumps}, we trigger an orbital jump, i.e., send the star back to the black hole at a separation $r_{\rm jump}$ by following the keplerian orbit analytically, without simulating the full orbital motion of the star around the black hole. In Tables \ref{first_passage} and \ref{last_passage},  all the models with an orbital jump are marked by a $\dagger$. Note that we list the results of the simulations for the models with $r_p=r_T$ and $M_{\rm BH} \geq 100\,M_{\odot}$ with and without an orbital jump in Table \ref{first_passage} to show that they lead to similar results. The small differences are due to the artificial viscosity being turned off for the runs with orbital jumps.

During an orbital jump, we treat the structure of the remnant as unchanged. To check if this approximation is robust, we estimate the thermal timescale throughout the disrupted star right before the orbital jump in several cases with large orbital periods. We use an approach similar to that of section 4.2.2 in \citet{Antonini_2011}, although we calculate a local timescale at each particle location.  Specifically, we evaluate the local thermal timescale as $u\rho/|\nabla \cdot {\mathbf F}|$, where $u$ is the specific internal energy density, $\rho$ is the density, and ${\mathbf F}$ is the radiative flux, calculated in the diffusion approximation.  We find that the mass affected by thermal relaxation on the timescale of the orbital period is not only small (typically $\lesssim 10^{-3} M_\odot$) but also smaller than the amount of mass stripped on the subsequent pericenter passage. Such comparisons provide strong evidence that the matter affected by thermal relaxation during the excised orbit would have been ultimately stripped even if thermal contraction had been modeled.

\begin{figure*}
    \centering
    \includegraphics[width=1\linewidth]{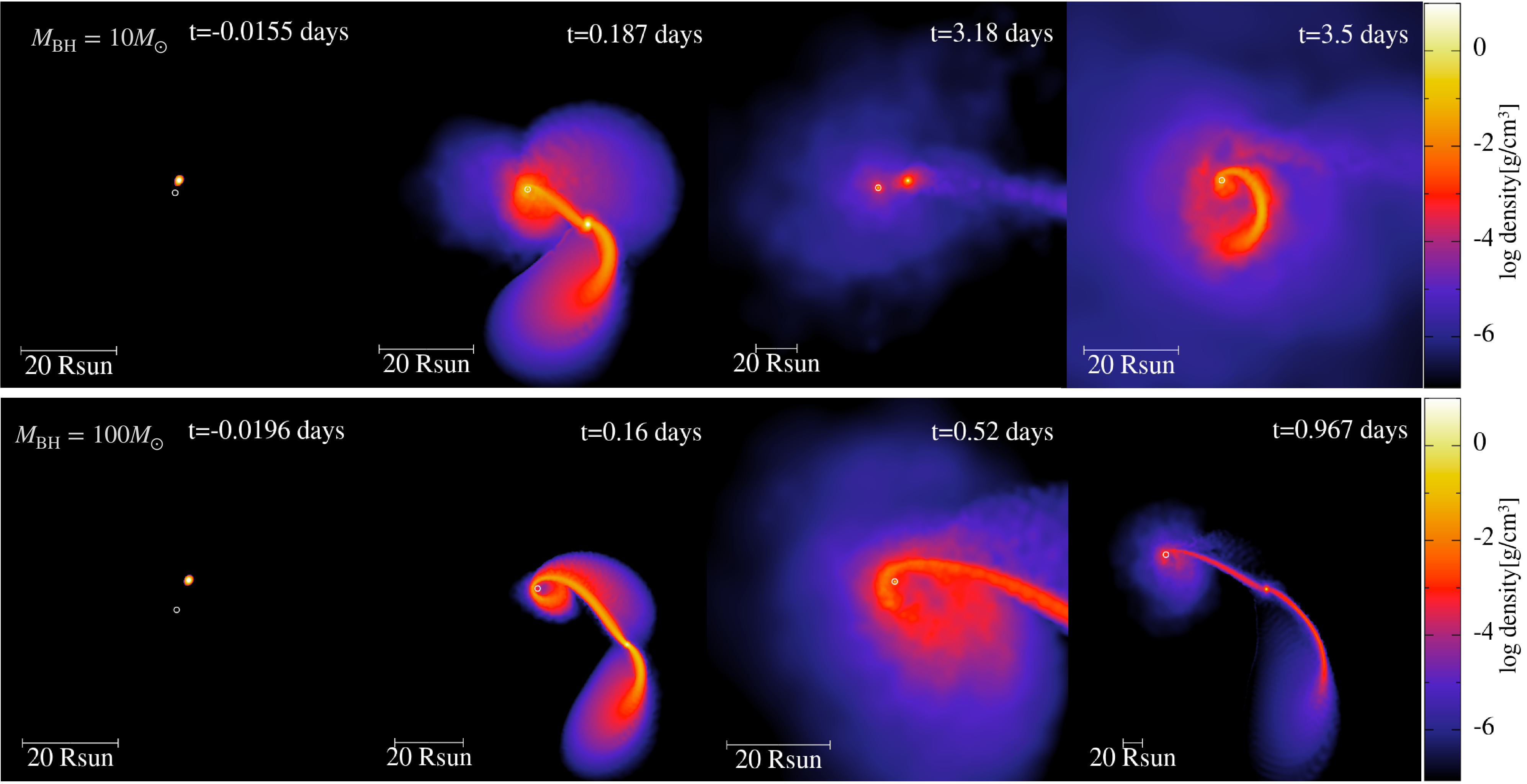}
    \caption{\footnotesize A $1\,M_{\odot}$ star encountering a $10\,M_{\odot}$ (above) and $100\,M_{\odot}$ (below) black hole with $r_p=0.4\,r_T$. Top: after the first pericenter passage,  the roughly $0.4\,M_{\odot}$ partially-disrupted stellar remnant becomes bound to the black hole. About 4 days after the first pericenter passage, the remnant returns to the pericenter and becomes disrupted completely by the $10\,M_{\odot}$ black hole. Bottom: after the first pericenter passage, the roughly $0.3\,M_{\odot}$ partially-disrupted stellar remnant is ejected with a kick velocity of roughly $300\,\rm{km\,s}^{-1}$.}\label{sph_plots}
    \label{fig:partial_unbound}
\end{figure*}

\begin{figure}
    \centering
    \includegraphics[width=\linewidth]{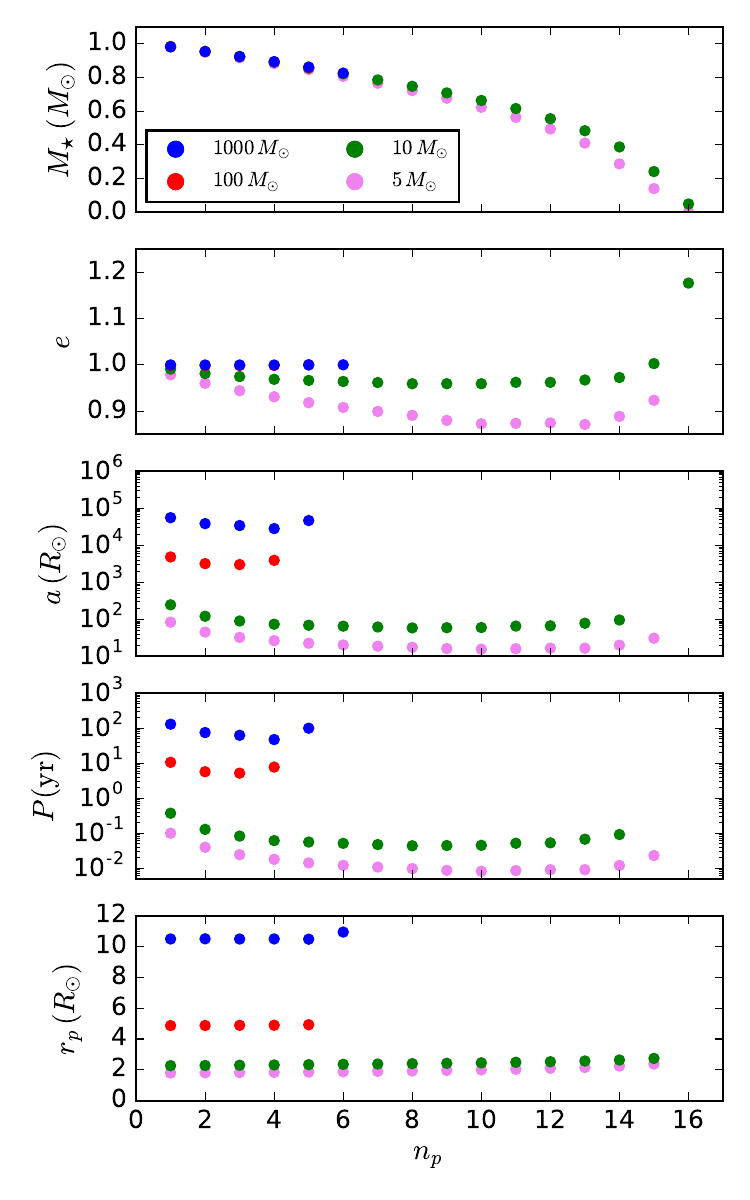}
    \caption{\footnotesize  Mass of the star $M_{\star}$, eccentricity $e$, semi-major axis $a$, period of the binary $P$, and the pericenter distance $r_p$ after each consequent pericenter passage. We show the pericenter passage number $n_p$ on the horizontal axis. Different black hole masses are shown with different colors (in all cases we assume a pericenter distance of $r_p=r_T$). After each pericenter passage, a small amount of material (that increases with each passage) is stripped from the star until, ultimately, the star is fully disrupted if $M_{\rm BH}=5\,M_{\odot}$ or ejected if $M_{\rm BH}\geq 10\,M_{\odot}$. The number of pericenter passages before ejection increases with black hole mass.}
    \label{fig:multiple_peri}
\end{figure}

In our recent study on the disruption of Eddington standard stellar models by stellar-mass black holes ($M_{\rm BH} \sim 10\,M_{\odot}$) \citep[][]{Kremer_2022}, we specifically show that full disruption is the final outcome only for nearly head-on collisions and/or mass ratios relatively close to unity. We show an example of this on the top panel of Figure \ref{sph_plots} for the $1\,M_{\odot}$ star encountering a $10\,M_{\odot}$ black hole with the pericenter distance $r_p=0.4\,r_T$. After the first pericenter passage, about $0.4\,M_{\odot}$ partially-disrupted stellar remnant becomes bound to the black hole. The remnant returns back to the pericenter after roughly 4 days and undergoes additional passage before being fully disrupted by the black hole. On the bottom figure, however, the $1\,M_{\odot}$ star interacting with an IMBH of mass $100\,M_{\odot}$ with the pericenter distance $r_p=0.4~r_T$ is unbound from the system after the first pericenter passage. In this case, roughly $0.3\,M_{\odot}$ partially-disrupted stellar remnant is ejected with a kick velocity of roughly $300\,\rm{km\,s}^{-1}$.

In Figure \ref{fig:multiple_peri}, we show the mass and the orbital properties of a $1\,M_{\odot}$ star that undergoes multiple pericenter passages after encountering black holes with various $M_{\rm BH}$. In each model, we assume a pericenter distance of $r_p/r_T =1$ and initial separation of $10\,r_T$, for which  we have $r_p^3/M_{\rm BH}=$constant and thus the same time to reach pericenter. Immediately after a jump occurs, however, the separation between the star and black hole is set to be the jump radius $r_{\rm jump}$, which turns out to be $7.9\,r_T$ for the $100\,M_{\odot}$ case and $17\,r_T$ for the $1000\,M_{\odot}$ case. Hence, after the first pericenter passage, the stars take a different amount of time to reach the next pericenter passage. In the case of mass ratios $M_{\star}/M_{\rm BH} \leq 0.1$, the stellar remnant is eventually ejected with more mass removed after each pericenter passage. Apart from this, the star is ultimately fully disrupted in the case of $M_{\rm BH}=5\,M_{\odot}$.

Additionally, the number of pericenter passages varies with the black hole mass. We show this trend in Figure \ref{fig:multiple_peri}, where the horizontal axis of each panel is the number of pericenter passages, $n_p$. As can be seen, lower black hole masses lead to a larger number of passages. For instance, at the pericenter distance $r_p=r_T$, the $1\,M_{\odot}$ star undergoes 16 (5) pericenter passages before ejection after encountering a black hole of mass $10\,M_{\odot}$ ($100\,M_{\odot}$). In the $100\,M_{\odot}$ case, the eccentricity is always larger than 0.999 and the semi-major axis is $\sim 10^4\,R_{\odot}$ before the ejection. Thus, the amount of mass loss necessary to eject the star in the final pericenter passage does not need to be a significant fraction of the initial stellar mass. In contrast, in the $10\,M_{\odot}$ case, the eccentricity is 0.96 and the semi-major axis is less than $100\,R_{\odot}$ right before the last pericenter passage. As the minimum orbital energy that would have to be injected to eject the star is $GM_{\rm BH}M_{\star}/(2a)$, the smaller semi-major axis in the $10\,M_{\odot}$ case requires at least 3 times higher mass loss and thus larger number of close passages to eject the star. In addition to the mass ratio, the specific number of close passages is determined by pericenter distance. We list the total number of pericenter passages, $n_p$, on Column 3 of Table \ref{last_passage}, which increases with the pericenter distance for each model.

\subsection{Electromagnetic signatures}

We also determine properties of the accretion disks formed by analyzing the structure of the material bound to the black holes (e.g., the disk) during our hydrodynamic simulations. 

We estimate the viscous accretion time scale for the stellar debris bound to the black hole as \citep[][]{ShakuraSunyaev1973}:
\begin{equation}
    t_{\rm acc} = [\alpha h^2 \Omega]^{-1},
    \label{eq:t_acc}
\end{equation}
where $\alpha$ is the dimensionless viscosity parameter (we assume $\alpha=0.1$),  $\Omega$ is the angular velocity of the disk, $h=H/R_{d}$ (where $H $ is the disk scale height and $R_{d}$ is the disk radius). In all calculations, we assume $h=1$ as in our previous study \citep{Kremer_2022}. This choice is motivated by previous simulations of super-Eddington disks that demonstrate the disks will rapidly ``puff up" and become thick as a result of disk-wind mass loss \citep[e.g.,][]{NarayanYi1995, Blandford1999, Yuan2012}.
It is not clear at what black hole mass the disk would transition from a thick disk to a more classic thin disk. We anticipate that much higher SMBH-like masses are required for this to matter and in our $M_{\rm BH} < 10^4 M_{\odot}$ case, the thick disk regime is mostly applicable since the mass transfer rates are highly super-Eddington.

In general, the viscous accretion time scale is longer than the typical fallback timescales. In our current SPH simulations, accretion processes are therefore not considered. However, we see the trend in our simulations that the fallback timescale becomes longer for very large black hole masses $M_{\rm BH}\gtrsim 10^4\,M_{\odot}$ and the TDE transitions from viscous driven to fallback driven as in the classic SMBH TDE case \citep[e.g.,][]{Rees_1988}.

In the case of multiple passages, we show that the total mass stripped from the star  (see Figure \ref{fig:multiple_peri}) and therefore the total bound mass to the black hole increases after each pericenter passage, suggesting an increase in the brightness of the repeated accretion flares with the final flare being the brightest. Observations of variable, repeated electromagnetic accretion flares could potentially indicate the presence of an accreting stellar black hole or IMBH.  One of the best candidates for an accreting IMBH is the variable X-ray source HLX-1 \citep[e.g.,][]{Farrell2009}, which has shown quasi-periodic outbursts with a period of $\sim 1$ yr. In the last several years, the time between outbursts has been increasing \citep[][]{2020MNRAS.491.5682L}, consistent with a mass-transfer system emerging from a minimum in orbital period, as shown in \cite{Godet_2014}. Indeed in Figure \ref{fig:multiple_peri}, we see that our $r_{\rm p}=r_{\rm T}$ calculations with $M_{\rm BH}=10\, M_\odot$ and $100 \,M_\odot$ straddle the observed period of HLX-1.  Among all our $r_p=r_T$ calculations, a black hole mass $M_{\rm BH}=50 \,M_\odot$ (case 39 in Tables \ref{first_passage} and \ref{last_passage}) comes closest to reproducing a $\sim 1$~yr period for a few orbits before the star is ejected. Spectral modeling of the observed X-ray and other wavelength data from HLX-1, however, implies that $M_{\rm BH}\gtrsim 10^4 M_\odot$ \citep[e.g.,][]{Straub_2014,2017MNRAS.469..886S} and so scenarios like the ones considered in this paper can be excluded as a possibility: if mass transfer from a donor on a highly eccentric orbit is the correct explanation of HLX-1, a white dwarf orbiting with larger $r_{\rm p}/r_{\rm T}$ is more likely \citep[][]{Godet_2014}.

\begin{figure}
    \centering
    \includegraphics[width=1\linewidth]{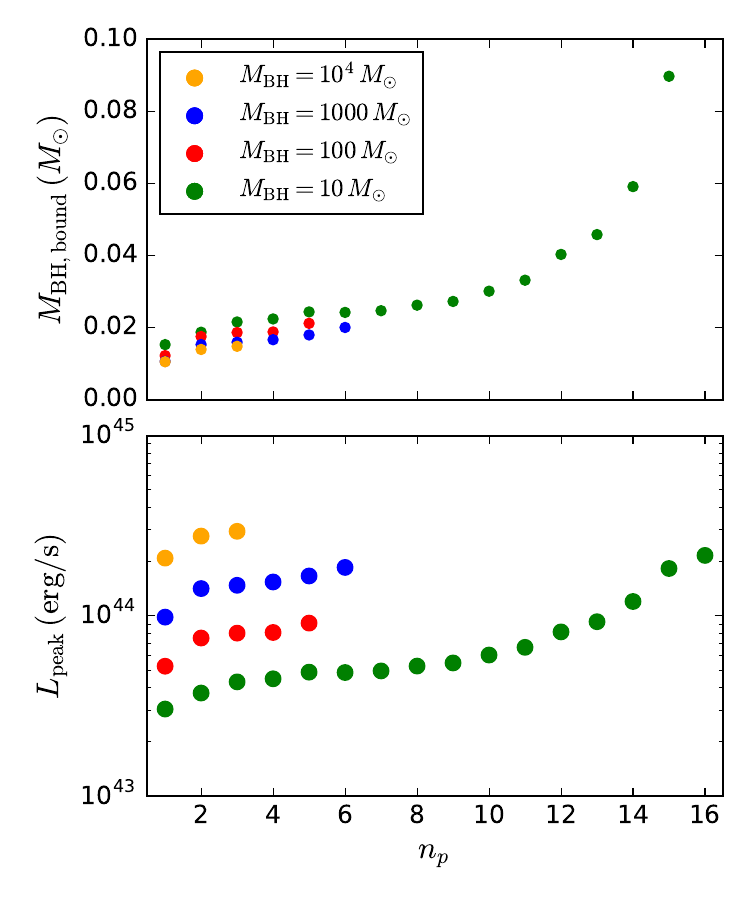}
    \caption{\footnotesize  Stripped material bound to the black hole $M_{\rm BH,bound}$, and peak luminosity from accretion onto black hole $L_{\rm peak}$ after each successive pericenter passage. We show the passage number $n_p$ on the horizontal axis. Peak luminosities are calculated right before an orbital jump is triggered and taking the value of the exponent to be $s=0.5$ in Equation \ref{eq:Lum}. }
    \label{fig:luminosity}
\end{figure}

The accretion rate onto the black hole is given by $\dot{M}\sim M_{\rm disk}/\Delta t$, where $M_{\rm disk}$ is the total disk mass and, the characteristic accretion timescale $\Delta t$ is the viscous accretion time of the disk, $t_{\rm{acc}}$ (Equation~\ref{eq:t_acc}).  
For $M_{\rm disk} \approx 10^{-3}-1\,M_{\odot}$ and $\Delta t\approx1-10\,$d, $\dot{M}$ ranges from roughly $10^{-2}-10^2\,M_{\odot}\,\rm{yr}^{-1}$, exceeding the classic Eddington accretion limit by several orders of magnitude, 
\begin{equation}
   \dot{M}_{\rm Edd} = \frac{L_{\rm Edd}}{\epsilon c^2}  \approx 2\times10^{-6} \left(\frac{M_{\rm BH}}{100\,M_{\odot}}\right) M_{\odot}\,\rm{yr}^{-1} 
\end{equation}
 adopting a radiative efficiency of 0.1 and taking $L_{\rm Edd} \approx 10^{40} (M_{\rm BH}/100\,M_{\odot}) \, \rm{erg\,s}^{-1}$ as the electron-scattering Eddington luminosity. In this ``hypercritical'' accretion regime, photons are trapped and the disk is unable to cool efficiently via radiation \citep[e.g.,][]{Begelman1979}, ultimately reducing the total mass supplied to the black hole via outflows and justifying our use of the thick disk ($h=1$) assumption.

The peak accretion luminosity from accretion onto the black hole can be estimated as
\begin{equation}
    \label{eq:Lum}
    L_{\rm{peak}} \approx \epsilon \dot{M}_{\rm{BH}} c^2 \approx \epsilon \frac{M_{\rm{disk}}}{\Delta t}\Bigg(\frac{R_{\rm in}}{R_d} \Bigg)^s c^2
\end{equation}
where $\epsilon$ is the accretion efficiency factor near the innermost stable circular orbit for which we assume a fiducial value of $10^{-2}$ \citep[][]{SadowskiNarayan2016}. For the material bound to the black hole and forms a disk, some will be accreted and some will be unbound from the system via a disk wind.
In line with previous studies of thick super-Eddington accretion disks  \citep[e.g.,][]{Blandford1999,NarayanYi1995}, here we assume that the accretion rate $\dot{M}_{\rm{BH}}$ is reduced by a factor of $(R_{\rm in}/R_d)^s$, where $R_{\rm in} = 6GM_{\rm BH}/c^2$ is the inner edge of the disk, $R_d$ is the outer edge of the disk, and the power-law index $s \in [0, 1]$ parameterizes the fraction of material transported from the outer edge of the disk to the black hole and determines the fraction of accreted versus wind-mass loss. In reality, a fraction of bound material is accreted ($<10\%$) and the remainder is blown away in a disk wind. The wind will be launched with a typical velocity $10^3-10^4~\rm{km/s}$ \citep[][]{Kremer_2019}. Given the typical orbital period between passages and the velocity of the wind, we estimate that the typical distance traveled by this wind is much larger than the TDE scale (of order a few stellar radii), allowing us to justify our assumption of removing all material before the next pericenter passage occurs.

As we consider black holes of larger mass, the orbit of the debris becomes increasingly eccentric, with $e_{\rm mb} = 1-r_p/a_{\rm mb}$, where $a_{\rm mb}=r_T^2/(2R_{\star})$ is the semi-major axis of the most bound material. For example, in our $r_p=r_T$ cases, the eccentricity of the most bound material is 0.07, 0.57, 0.80, and 0.91 for black hole masses 10, 100, 1000, $10^4\,M_{\odot}$, respectively. Indeed, calculating the debris structure’s radial size directly from SPH simulations as in \cite{Kremer_2022} gives, for $M_{\rm BH} \gtrsim 10^3\,M_{\odot}$, a disk radius much larger than the innermost edge of the disk ($\sim r_p$). This would potentially result in accretion timescales that are hundreds of times longer than the orbital period of the most bound stellar debris \citep{Canizzo_1990,Dai_2015} $P_{\rm mb}=2\pi \sqrt{a_{\rm mb}^3/(GM_{\rm BH})}\approx 0.004\,{\rm yr}\sqrt{M_{\rm BH}/(10^3 M_\odot)}$ for our $1 M_\odot$ star. The period plot on the fourth panel of Figure \ref{fig:multiple_peri} gives us an estimate for the time between accretion flares that is much longer than the circularization timescale ($\sim 100\,P_{\rm mb}$). Thus, we make the assumption that the disk is circularized before next pericenter passage by taking the disk radius $R_d=2r_p$. 

In Figure \ref{fig:luminosity}, we show the stripped material bound to the black hole $M_{\rm BH,bound}$, and the peak accretion luminosity $L_{\rm peak}$ using Equation (\ref{eq:Lum}) right before an orbital jump is triggered after each pericenter passage. Different colors indicate different black hole masses and the pericenter distance is set to $r_p=r_T$ in all models. For each model, we use the viscous accretion timescale taking the circularized disk radius to be $R_d=2r_p$ and the angular velocity to be keplerian ($\Omega \propto r^{-3/2}$). Because the estimated accretion rates are highly super-Eddington, we adopt the thick disk approximation ($H/R_{\rm d}\approx1$) in line with previous studies of super-Eddington disks \citep[e.g.,][]{Blandford1999}. We consider as perhaps the most likely realistic case using the exponent $s=0.5$ in Equation (\ref{eq:Lum}) based on numerical simulations of adiabatic accretion flows of \cite{Yuan2012, Yuan2014}. 
 In the case of $M_{\rm BH}=10~M_{\odot}$, the star undergoes 16 pericenter passages in total before it is ejected with a kick velocity of $v_{\rm kick, \star} \approx 300\,\rm{km/s}$.  
We see that successive accretion flares exhibit an increase in brightness with the final flare being the brightest, e.g., the peak luminosity increases by several factors from the first pericenter passages to the last, exceeding $L_{\rm peak}\sim 10^{44}~\rm{erg/s}$. 

\section{Discussion $\&$ Conclusions} \label{discussion}

In this paper, we have explored the tidal disruption of a $1\,M_{\odot}$ MS star by massive black holes by performing a suite of SPH calculations. We determine the boundary between complete and partial disruptions by examining a large set of results for different pericenter distances and black hole masses. We find that for the strongest encounters ($r_p < 0.3\,r_T$), the star is disrupted fully after the first pericenter passage. For weaker encounters, the star is partially disrupted with its dense core surviving. As in previous studies, we show that the mass of the stellar remnant can be simply estimated for a given pericenter distance using the density profile of the star and a numerical factor, which depends on the central concentration. For more massive black holes, the remnant star becomes unbound for a wider range of $r_p/r_T$ after receiving an impulsive kick.

Here we find that the kick velocity the star receives \textit{increases} with increasing black hole mass and also increases as more mass is stripped at the pericenter. The hydrodynamic simulations of TDEs around SMBHs \citep[][]{Manukian_2013,Gafton_2015}, on the other hand, indicate that the kick velocity is independent of the mass ratio but increases with the mass loss. In the IMBH regime, we see that kick velocities increase with black hole mass even though the mass stripped from the star remains roughly the same at a given $\beta$. This is because a larger kick can still be imparted to the star due to the larger ejection speed of the stripped material. Our calculations show that the kick velocities can be as high as $10^3\,\rm{km/s}$ when $M_{\rm BH}\gtrsim 10^3\,M_{\odot}$. Indeed, it has been speculated that some hypervelocity stars could originate from encounters with an IMBH in young star clusters. The best candidate is the hypervelocity star HE 0437–5439, which could have been ejected from the Large Magellanic Cloud \citep[][]{Edelmann_2005} after interacting with an IMBH more massive than about $1000\,M_{\odot}$ \citep[][]{Gualandris_2007}.

The multiple passages are one of the key results of this paper. For weaker encounters, the partially disrupted star is tidally captured by the black hole with the total number of subsequent close passages depending on the mass ratio and pericenter distance. For an encounter with a stellar-mass black hole, the star returns to the pericenter for one or more subsequent pericenter passages until ultimately being disrupted fully or ejected. On the other hand, in all encounters with an IMBH and $r_p/r_T> 0.3$, the remnant star is ultimately ejected after the last pericenter passage. Additionally, we find that for a fixed penetration factor $r_T/r_p$, interactions with a more massive black hole lead to fewer pericenter passages before the star is ejected. As each of these close passages is expected to produce a brief electromagnetic flare from accretion, one could in principle constrain the black hole mass from the number and properties of flares observed. However, the degeneracy between the $r_p/r_T$ value and the black hole mass in terms of the total number of successive close passages could make this difficult.

 We also estimate the properties of the accretion flares from successive strippings, including flare luminosities and time between flares. The partial disruption of a $1\,M_{\odot}$ star yields a minimum orbital period a few $\times 10\,(10^3)\,$yr around a $100\,M_{\odot}\,(10^4\,M_{\odot})$ IMBH. However, one can still obtain a partially disrupted star on an orbit around a massive black hole as short as a few years through a dynamical exchange of the star that was initially part of a binary system \citep{Cufari_2022,Wevers_2023}. Recently, \cite{Wevers_2023} proposed that the TDE AT 2018fyk is caused by repeating partial disruption of a star through the Hills mechanism and predicted the orbital period of the captured star around $10^{7.7}\,M_{\odot}$ black hole to be $\sim 1200$ days. 
 
We find that the brightness of successive accretion flares increases after each pericenter passage as the amount of the stripped material bound to the black hole increase. For example, with an IMBH of mass $M_{\rm BH} = 100\,M_{\odot}$ and a $1\,M_{\odot}$ mass star initially at $r_p=r_T$, the peak luminosity increases by several factors from the first pericenter passage to the last, when it reaches $10^{44}~$erg/s.
Our calculations for peak luminosities are roughly in agreement with the estimate by \cite{Chen_2018} who found that a long-term ($\sim 10$ yr) and luminous ($\sim 10^{42}~$erg/s) emission during the super-Eddington accretion phase following a TDE could hint at the presence of IMBHs in GCs and dwarf galaxies.

There are several aspects of these processes that have not yet been taken into account in our work. Obviously, the parameter space for these interactions is considerably larger and one could, for example, consider stars of different masses and evolutionary stages. Throughout this work, we make the simple assumption that the debris stream circularizes and settles down to a disk very rapidly after returning back to the black hole. However, the gas must lose an amount of energy $\sim G M_{\rm BH}/R$ per unit mass at radius $R$ to circularize the falling matter \citep{Piran_2015}. In an IMBH TDE, it has been suggested that energy dissipation is in fact much slower, making the circularization of the debris stream very inefficient \citep{Shiokawa_2015,Chen_2018}. In that case, much of the stellar debris is held in an elliptical disk for a long time \citep[$\sim 10\,$years;][]{Ramirez_Ruiz_2009,Guillochon_2014,Chen_2018}, resulting in accretion timescales of hundreds of times longer than the orbital period of the most bound stellar debris \citep{Canizzo_1990,Dai_2015}. In this case, the peak luminosities estimated in Figure~\ref{fig:luminosity} may overestimate the true peak luminosities by up to a factor of roughly $10^2-10^3$, specifically for the most massive black holes which produce the most elliptical disks and therefore have the longest circularization timescales. Future work should therefore address the circularization for these events and have better estimates for the peak luminosities. Additionally, future works should attempt to include self-consistently radiation and accretion feedback processes on the hydrodynamics. With a large database of SPH results, one could then derive fitting formulae for the outcomes of all encounters between black holes and stars to include into $N$-body simulations of dense star clusters.
Finally, our current models assume parabolic encounters, suitable for star clusters with low-velocity dispersions, such as GCs, but studying TDEs and collisions in more massive star clusters, such as nuclear star clusters, will require calculations for hyperbolic encounters.

\begin{acknowledgements}

	We would like to thank an anonymous referee for insightful suggestions. We also thank Elena Gonz{\'{a}}lez for useful discussions. FK acknowledges support from a CIERA Board of Visitors Graduate Fellowship. KK is supported by an NSF Astronomy and Astrophysics Postdoctoral Fellowship under the award AST-2001751. GF acknowledges support from NASA Grant 80NSSC21K1722. This work was supported by NSF Grant AST-2108624 and NASA ATP Grant 80NSSC22K0722 at Northwestern University. We thank Byron Rich for helping to prepare workstations at Allegheny College, purchased with a grant from the George I.$\,$Alden Trust, to run several of the simulations of this paper.
\end{acknowledgements}
\appendix

We include in the Appendix two tables containing detailed information for each simulation. In Table \ref{first_passage} and  \ref{last_passage}, we list simulation outcomes after the first and last  pericenter passage, respectively.

\startlongtable
\begin{deluxetable*}{l|rcc|cccc|cccccc|c}
\tabletypesize{\footnotesize}
\tablewidth{0pt}
\tablecaption{Results after first pericenter passage}
\label{first_passage}
\tablehead{
	\colhead{} &
	\colhead{$^1M_{\rm{BH}}$} &
	\colhead{$^2r_{\rm p}/r_{\rm T}$} &
	\colhead{$^3$outcome} &
	\colhead{$^4M_{\rm{bound,BH}}$} &
	\colhead{$^5M_{\rm{ej}}$} &
	\colhead{$^6M_{\star,f}$} &
	\colhead{$^7R_{\star,99}$} &
	\colhead{$^8P_{\rm{orb}}$} &
	\colhead{$^{9}v_{\rm{kick,*}}$} &
	\colhead{$^{10}v_{\rm{kick,BH}}$} \\
	\colhead{} &
	\colhead{$M_{\odot}$} &
	\colhead{} &
    \colhead{} &
	\colhead{$M_{\odot}$} &
	\colhead{$M_{\odot}$} &
	\colhead{$M_{\odot}$} &
	\colhead{$R_{\odot}$} &
	\colhead{days} &
	\colhead{km\,s$^{-1}$} &
	\colhead{km\,s$^{-1}$}
}
\startdata
1$^\dagger$ & 5     & 1.00 & B & 0.017 & 0.001 & 0.982 & 0.893    & 38.3     &0.0827   & 0.0827   \\ % rt_Mbh5_Mstar1_v0_eqmass0_N1e5_noAV_rjump
\hline
2           & 10    & 0.20 & F & 0.672 & 0.329 & 0.000 & 0        & N/A & N/A & 18.1     \\ % r02t_Mbh10_Mstar1_v0_eqmass0_N1e5/ 
3$^\dagger$ & 10    & 0.30 & B & 0.407 & 0.270 & 0.323 & 2.83     & 1.26     &21.4     & 21.4     \\ % jump/r03t_Mbh10_Mstar1_v0_eqmass0_N1e5_noAV_rjump/ 
4$^\dagger$ & 10    & 0.43 & B & 0.333 & 0.178 & 0.489 & 2.24     & 4.05     &15.5     & 15.5     \\ % jump/r04t_Mbh10_Mstar1_v0_eqmass0_N1e5_noAV_rjump/ 
5$^\dagger$ & 10    & 0.50 & B & 0.270 & 0.124 & 0.606 & 2.32     & 6.13     &11       & 11       \\ % jump/r05t_Mbh10_Mstar1_v0_eqmass0_N1e5_noAV_rjump/ 
6$^\dagger$ & 10    & 0.60 & B & 0.179 & 0.063 & 0.758 & 2.45     & 9.71     &5.54     & 5.54     \\ % jump/r06t_Mbh10_Mstar1_v0_eqmass0_N1e5_noAV_rjump/ 
7$^\dagger$ & 10    & 0.70 & B & 0.106 & 0.028 & 0.866 & 2.35     & 15.3     &2.42     & 2.42     \\ % jump/r07t_Mbh10_Mstar1_v0_eqmass0_N1e5_noAV_rjump/ 
8$^\dagger$ & 10    & 0.85 & B & 0.043 & 0.007 & 0.950 & 1.60     & 39.2     &0.614    & 0.614    \\ % jump/r085t_Mbh10_Mstar1_v0_eqmass0_N1e5_noAV_rjump/ 
9$^\dagger$ & 10    & 1.00 & B & 0.016 & 0.002 & 0.982 & 0.828    & 138      &0.127    & 0.127    \\ % jump/rt_Mbh10_Mstar1_v0_eqmass0_N1e5_noAV_rjump/ 
10$^\dagger$ & 10    & 1.25 & B & 0.003 & 0.000 & 0.997 & 0.653    & 1.25e+03 &0.00484  & 0.00484  \\ % jump/r1.25t_Mbh10_Mstar1_v0_eqmass0_N1e5_noAV_rjump/ 
11$^\dagger$ & 10    & 1.50 & B & 0.001 & 0.000 & 1.000 & 0.645    & 9.35e+03 &2.97e-05 & 2.97e-05 \\ % jump/r1.5t_Mbh10_Mstar1_v0_eqmass0_N1e5_noAV_rjump/ 
\hline
12           & 20    & 0.20 & F & 0.554 & 0.446 & 0.000 & 0        & N/A & N/A & 15.2     \\ % jump/r02t_Mbh20_Mstar1_v0_eqmass0_N1e5/ 
13$^\dagger$ & 20    & 0.30 & B & 0.488 & 0.294 & 0.218 & 2.36     & 568      &15.4     & 15.4     \\ % jump/r03t_Mbh20_Mstar1_v0_eqmass0_N1e5_noAV_rjump/ 
14$^\dagger$ & 20    & 0.40 & B & 0.402 & 0.240 & 0.358 & 2.45     & 358      &13.1     & 13.1     \\ % jump/r04t_Mbh20_Mstar1_v0_eqmass0_N1e5_noAV_rjump/ 
15$^\dagger$ & 20    & 0.50 & B & 0.294 & 0.160 & 0.546 & 2.84     & 64.7     &8.79     & 8.79     \\ % jump/r05t_Mbh20_Mstar1_v0_eqmass0_N1e5_noAV_rjump/ 
16$^\dagger$ & 20    & 0.60 & B & 0.187 & 0.086 & 0.727 & 2.98     & 56.6     &4.69     & 4.69     \\ % jump/r06t_Mbh20_Mstar1_v0_eqmass0_N1e5_noAV_rjump/ 
17$^\dagger$ & 20    & 0.70 & B & 0.107 & 0.040 & 0.853 & 2.77     & 63.9     &2.12     & 2.12     \\ % jump/r07t_Mbh20_Mstar1_v0_eqmass0_N1e5_noAV_rjump/ 
18$^\dagger$ & 20    & 0.85 & B & 0.040 & 0.011 & 0.949 & 1.53     & 133      &0.561    & 0.561    \\ % jump/r085t_Mbh20_Mstar1_v0_eqmass0_N1e5_noAV_rjump/ 
19$^\dagger$ & 20    & 1.00 & B & 0.015 & 0.003 & 0.982 & 0.799    & 435      &0.135    & 0.135    \\ % jump/rt_Mbh20_Mstar1_v0_eqmass0_N1e5_noAV_rjump/ 
\hline
20           & 30    & 0.20 & F & 0.505 & 0.495 & 0.000 & 0        & N/A & N/A & 13       \\ % r02t_Mbh30_Mstar1_v0_eqmass0_N1e5/ 
21           & 30    & 0.30 & U & 0.494 & 0.327 & 0.179 & 2.46     & N/A & 201      & 13.6     \\ % jump/r03t_Mbh30_Mstar1_v0_eqmass0_N1e5/ 
22           & 30    & 0.40 & U & 0.411 & 0.269 & 0.320 & 2.74     & N/A & 132      & 11.9     \\ % jump/r04t_Mbh30_Mstar1_v0_eqmass0_N1e5/ 
23$^\dagger$ & 30    & 0.50 & B & 0.302 & 0.180 & 0.518 & 3.23     & 1.74e+05 &7.25     & 7.25     \\ % jump/r05t_Mbh30_Mstar1_v0_eqmass0_N1e5_noAV_rjump/ 
24$^\dagger$ & 30    & 0.60 & B & 0.188 & 0.099 & 0.713 & 3.39     & 197      &3.94     & 3.94     \\ % jump/r06t_Mbh30_Mstar1_v0_eqmass0_N1e5_noAV_rjump/ 
25$^\dagger$ & 30    & 0.70 & B & 0.105 & 0.046 & 0.849 & 2.95     & 150      &1.78     & 1.78     \\ % jump/r07t_Mbh30_Mstar1_v0_eqmass0_N1e5_noAV_rjump/ 
26$^\dagger$ & 30    & 0.85 & B & 0.039 & 0.013 & 0.947 & 1.54     & 252      &0.494    & 0.494    \\ % jump/r085t_Mbh30_Mstar1_v0_eqmass0_N1e5_noAV_rjump/ 
27$^\dagger$ & 30    & 1.00 & B & 0.014 & 0.003 & 0.982 & 0.777    & 799      &0.126    & 0.126    \\ % jump/rt_Mbh30_Mstar1_v0_eqmass0_N1e5_noAV_rjump/ 
\hline
28           & 50    & 0.20 & F & 0.457 & 0.543 & 0.000 & 0        & N/A & N/A & 10       \\ % r02t_Mbh50_Mstar1_v0_eqmass0_N1e5/ 
29           & 50    & 0.30 & U & 0.514 & 0.355 & 0.130 & 2.44     & N/A & 359      & 10.2     \\ % r03t_Mbh50_Mstar1_v0_eqmass0_N1e5/ 
30           & 50    & 0.40 & U & 0.428 & 0.300 & 0.272 & 2.99     & N/A & 243      & 9.23     \\ % r04t_Mbh50_Mstar1_v0_eqmass0_N1e5/ 
31           & 50    & 0.50 & U & 0.299 & 0.199 & 0.502 & 3.85     & N/A & 101      & 6.29     \\ % r05t_Mbh50_Mstar1_v0_eqmass0_N1e5/ 
32$^\dagger$ & 50    & 0.60 & B & 0.188 & 0.111 & 0.701 & 3.90     & 3.19e+03 &2.98     & 2.98     \\ % jump/r06t_Mbh50_Mstar1_v0_eqmass0_N1e5_noAV_rjump/ 
33$^\dagger$ & 50    & 0.70 & B & 0.104 & 0.053 & 0.844 & 3.11     & 456      &1.38     & 1.38     \\ % jump/r07t_Mbh50_Mstar1_v0_eqmass0_N1e5_noAV_rjump/ 
34$^\dagger$ & 50    & 0.85 & B & 0.038 & 0.016 & 0.947 & 1.46     & 539      &0.391    & 0.391    \\ % jump/r085t_Mbh50_Mstar1_v0_eqmass0_N1e5_noAV_rjump/ 
35$^\dagger$ & 50    & 1.00 & B & 0.013 & 0.004 & 0.983 & 0.769    & 1.63e+03 &0.103    & 0.103    \\ % jump/rt_Mbh50_Mstar1_v0_eqmass0_N1e5_noAV_rjump/ 
36$^\dagger$ & 50    & 1.25 & B & 0.002 & 0.000 & 0.997 & 0.659    & 1.31e+04 &0.00782  & 0.00782  \\ % jump/r1.25t_Mbh50_Mstar1_v0_eqmass0_N1e5_noAV_rjump/ 
37$^\dagger$ & 50    & 1.50 & B & 0.000 & 0.000 & 1.000 & 0.649    & 7.26e+04 &0.000125 & 0.000125 \\ % jump/r1.5t_Mbh50_Mstar1_v0_eqmass0_N1e5_noAV_rjump/ 
\hline
38           & 100   & 0.22 & F & 0.423 & 0.577 & 0.000 & 0        & N/A & N/A & 6.22     \\ % r1_Mbh100_Mstar1_v0_eqmass0_N1e5/ 
39           & 100   & 0.32 & U & 0.525 & 0.375 & 0.100 & 2.40     & N/A & 568      & 6.27     \\ % r1.6_Mbh100_Mstar1_v0_eqmass0_N1e5/ 
40           & 100   & 0.43 & U & 0.397 & 0.300 & 0.303 & 4.09     & N/A & 267      & 5.37     \\ % r2_Mbh100_Mstar1_v0_eqmass0_N1e5/ 
41           & 100   & 0.54 & U & 0.251 & 0.178 & 0.571 & 4.78     & N/A & 125      & 3.42     \\ % r2.5_Mbh100_Mstar1_v0_eqmass0_N1e5/ 
42           & 100   & 0.60 & U & 0.181 & 0.120 & 0.698 & 4.50     & N/A & 40.3     & 2.11     \\ % r2.9_Mbh100_Mstar1_v0_eqmass0_N1e5/ 
43$^\dagger$ & 100   & 0.65 & B & 0.138 & 0.087 & 0.775 & 4.03     & 3.39e+04 &1.33     & 1.33     \\ % jump/r0.65t_Mbh100_Mstar1_v0_eqmass0_N1e5_noAV_rjump/ 
44$^\dagger$ & 100   & 0.70 & B & 0.101 & 0.060 & 0.839 & 3.43     & 2.06e+03 &0.907    & 0.907    \\ % jump/r0.7t_Mbh100_Mstar1_v0_eqmass0_N1e5_noAV_rjump/ 
45$^\dagger$ & 100   & 0.85 & B & 0.036 & 0.018 & 0.946 & 1.47     & 1.4e+03  &0.262    & 0.262    \\ % jump/r0.85t_Mbh100_Mstar1_v0_eqmass0_N1e5_noAV_rjump/ 
46           & 100   & 1.00 & B & 0.012 & 0.005 & 0.983 & 0.75     & 3.94e+03 &0.0681   & 0.0681   \\ % rt_Mbh100_Mstar1_v0_eqmass0_N1e5/ 
47$^\dagger$ & 100   & 1.00 & B & 0.012 & 0.005 & 0.982 & 0.779    & 3.91e+03 &0.0717   & 0.0717   \\ % jump/rt_Mbh100_Mstar1_v0_eqmass0_N1e5_noAV_rjump/ 
48$^\dagger$ & 100   & 1.10 & B & 0.006 & 0.002 & 0.992 & 0.695    & 9.31e+03 &0.0276   & 0.0276   \\ % jump/r1.10t_Mbh100_Mstar1_v0_eqmass0_N1e5_noAV_rjump/ 
49$^\dagger$ & 100   & 1.25 & B & 0.002 & 0.000 & 0.998 & 0.663    & 2.98e+04 &0.00613  & 0.00613  \\ % jump/r1.25t_Mbh100_Mstar1_v0_eqmass0_N1e5_noAV_rjump/ 
50$^\dagger$ & 100   & 1.50 & B & 0.000 & 0.000 & 1.000 & 0.650    & 1.52e+05 &0.000148 & 0.000148 \\ % jump/r1.5t_Mbh100_Mstar1_v0_eqmass0_N1e5_noAV_rjump/ 
51$^\dagger$ & 100   & 1.75 & B & 0.000 & 0.000 & 1.000 & 0.646    & 9.53e+05 &2.46e-08 & 2.46e-08 \\ % jump/r1.75t_Mbh100_Mstar1_v0_eqmass0_N1e5_noAV_rjump/ 
\hline
52           & 200   & 0.17 & F & 0.384 & 0.616 & 0.000 & 0        & N/A & N/A & 3.71     \\ % r1_Mbh200_Mstar1_v0_eqmass0_N1e5_bettercns/ 
53           & 200   & 0.32 & U & 0.516 & 0.417 & 0.066 & 2.63     & N/A & 580      & 3.79     \\ % r1.9_Mbh200_Mstar1_v0_eqmass0_N1e5/ 
54           & 200   & 0.43 & U & 0.407 & 0.327 & 0.266 & 4.90     & N/A & 312      & 3.26     \\ % r2.5_Mbh200_Mstar1_v0_eqmass0_N1e5/ 
55           & 200   & 0.51 & U & 0.283 & 0.220 & 0.497 & 5.61     & N/A & 183      & 2.38     \\ % r3_Mbh200_Mstar1_v0_eqmass0_N1e5/ 
56           & 200   & 0.60 & U & 0.178 & 0.129 & 0.693 & 4.91     & N/A & 76.1     & 1.38     \\ % r3.5_Mbh200_Mstar1_v0_eqmass0_N1e5/ 
57$^\dagger$ & 200   & 0.65 & U & 0.136 & 0.095 & 0.769 & 4.51     & N/A & 52.6     & 1.01     \\ % jump/r0.65t_Mbh200_Mstar1_v0_eqmass0_N1e5_noAV_rjump/ 
58$^\dagger$ & 200   & 0.70 & B & 0.098 & 0.065 & 0.837 & 3.79     & 8.43e+03 &0.562    & 0.562    \\ % jump/r0.7t_Mbh200_Mstar1_v0_eqmass0_N1e5_noAV_rjump/ 
59$^\dagger$ & 200   & 0.85 & B & 0.034 & 0.020 & 0.945 & 1.59     & 3.33e+03 &0.167    & 0.167    \\ % jump/r0.85t_Mbh200_Mstar1_v0_eqmass0_N1e5_noAV_rjump/ 
60           & 200   & 1.00 & B & 0.011 & 0.006 & 0.983 & 0.747    & 8.91e+03 &0.044    & 0.044    \\ % rt_Mbh200_Mstar1_v0_eqmass0_N1e5/ 
61$^\dagger$ & 200   & 1.00 & B & 0.012 & 0.006 & 0.982 & 0.801    & 8.83e+03 &0.0469   & 0.0469   \\ % jump/rt_Mbh200_Mstar1_v0_eqmass0_N1e5_noAV_rjump/ 
62$^\dagger$ & 200   & 1.10 & B & 0.006 & 0.002 & 0.992 & 0.704    & 2.05e+04 &0.0187   & 0.0187   \\ % jump/r1.10t_Mbh200_Mstar1_v0_eqmass0_N1e5_noAV_rjump/ 
63$^\dagger$ & 200   & 1.25 & B & 0.002 & 0.001 & 0.998 & 0.667    & 6.45e+04 &0.00436  & 0.00436  \\ % jump/r1.25t_Mbh200_Mstar1_v0_eqmass0_N1e5_noAV_rjump/ 
64$^\dagger$ & 200   & 1.50 & B & 0.000 & 0.000 & 1.000 & 0.650    & 3.09e+05 &0.000136 & 0.000136 \\ % jump/r1.5t_Mbh200_Mstar1_v0_eqmass0_N1e5_noAV_rjump/ 
65$^\dagger$ & 200   & 1.75 & B & 0.000 & 0.000 & 1.000 & 0.646    & 1.97e+06 &1.5e-08  & 1.5e-08  \\ % jump/r1.75t_Mbh200_Mstar1_v0_eqmass0_N1e5_noAV_rjump/ 
\hline
66           & 500   & 0.25 & F & 0.362 & 0.638 & 0.000 & 0        & N/A & N/A & 1.79     \\ % r2_Mbh500_Mstar1_v0_eqmass0_N1e5/ 
67           & 500   & 0.32 & U & 0.506 & 0.455 & 0.039 & 3.09     & N/A & 534      & 1.87     \\ % r2.5_Mbh500_Mstar1_v0_eqmass0_N1e5/ 
68           & 500   & 0.38 & U & 0.488 & 0.421 & 0.092 & 4.21     & N/A & 459      & 1.78     \\ % r3_Mbh500_Mstar1_v0_eqmass0_N1e5/ 
69           & 500   & 0.50 & U & 0.293 & 0.245 & 0.462 & 6.74     & N/A & 214      & 1.22     \\ % r4_Mbh500_Mstar1_v0_eqmass0_N1e5/ 
70           & 500   & 0.63 & U & 0.144 & 0.112 & 0.744 & 3.69     & N/A & 58.9     & 0.549    \\ % r5_Mbh500_Mstar1_v0_eqmass0_N1e5/ 
71$^\dagger$ & 500   & 0.65 & U & 0.133 & 0.103 & 0.763 & 5.18     & N/A & 66.6     & 0.518    \\ % jump/r0.65t_Mbh500_Mstar1_v0_eqmass0_N1e5_noAV_rjump/ 
72$^\dagger$ & 500   & 0.70 & B & 0.095 & 0.072 & 0.833 & 4.21     & 6.63e+04 &0.29     & 0.29     \\ % jump/r0.7t_Mbh500_Mstar1_v0_eqmass0_N1e5_noAV_rjump/ 
73$^\dagger$ & 500   & 0.85 & B & 0.033 & 0.023 & 0.944 & 1.62     & 9.44e+03 &0.0882   & 0.0882   \\ % jump/r0.85t_Mbh500_Mstar1_v0_eqmass0_N1e5_noAV_rjump/ 
74           & 500   & 1.00 & B & 0.010 & 0.006 & 0.983 & 0.742    & 2.42e+04 &0.0234   & 0.0234   \\ % rt_Mbh500_Mstar1_v0_eqmass0_N1e5/ 
75$^\dagger$ & 500   & 1.00 & B & 0.011 & 0.007 & 0.982 & 0.817    & 2.38e+04 &0.0255   & 0.0255   \\ % jump/rt_Mbh500_Mstar1_v0_eqmass0_N1e5_noAV_rjump/ 
76$^\dagger$ & 500   & 1.10 & B & 0.005 & 0.003 & 0.992 & 0.714    & 5.47e+04 &0.0104   & 0.0104   \\ % jump/r1.10t_Mbh500_Mstar1_v0_eqmass0_N1e5_noAV_rjump/ 
\hline
77           & 1000  & 0.20 & F & 0.469 & 0.531 & 0.000 & 0        & N/A & N/A & 1.09     \\ % r2_Mbh1000_Mstar1_v0_eqmass0_N1e5_c1_0.02/ 
78           & 1000  & 0.30 & U & 0.491 & 0.482 & 0.027 & 3.60     & N/A & 938      & 1.08     \\ % r3_Mbh1000_Mstar1_v0_eqmass0_N1e5/ 
79           & 1000  & 0.40 & U & 0.446 & 0.407 & 0.147 & 6.95     & N/A & 392      & 0.988    \\ % r4_Mbh1000_Mstar1_v0_eqmass0_N1e5/ 
80           & 1000  & 0.50 & U & 0.295 & 0.258 & 0.447 & 7.53     & N/A & 226      & 0.709    \\ % r5_Mbh1000_Mstar1_v0_eqmass0_N1e5/ 
81           & 1000  & 0.60 & U & 0.171 & 0.142 & 0.687 & 4.42     & N/A & 99.4     & 0.402    \\ % r6_Mbh1000_Mstar1_v0_eqmass0_N1e5/ 
82$^\dagger$ & 1000  & 0.65 & U & 0.131 & 0.109 & 0.761 & 5.58     & N/A & 58.3     & 0.292    \\ % jump/r0.65t_Mbh1000_Mstar1_v0_eqmass0_N1e5_noAV_rjump/ 
83$^\dagger$ & 1000  & 0.70 & B & 0.095 & 0.075 & 0.830 & 4.30     & 4.01e+05 &0.172    & 0.172    \\ % jump/r0.7t_Mbh1000_Mstar1_v0_eqmass0_N1e5_noAV_rjump/ 
84$^\dagger$ & 1000  & 0.85 & B & 0.032 & 0.024 & 0.943 & 1.69     & 1.89e+04 &0.053    & 0.053    \\ % jump/r0.85t_Mbh1000_Mstar1_v0_eqmass0_N1e5_noAV_rjump/ 
85           & 1000  & 1.00 & B & 0.010 & 0.007 & 0.983 & 0.738    & 4.95e+04 &0.0142   & 0.0142   \\ % rt_Mbh1000_Mstar1_v0_eqmass0_N1e5/ 
86$^\dagger$ & 1000  & 1.00 & B & 0.011 & 0.008 & 0.982 & 0.813    & 4.85e+04 &0.0157   & 0.0157   \\ % jump/rt_Mbh1000_Mstar1_v0_eqmass0_N1e5_noAV_rjump/ 
87$^\dagger$ & 1000  & 1.10 & B & 0.005 & 0.003 & 0.992 & 0.714    & 1.12e+05 &0.00651  & 0.00651  \\ % jump/r1.10t_Mbh1000_Mstar1_v0_eqmass0_N1e5_noAV_rjump/ 
\hline
88           & 10000 & 0.20 & F & 0.509 & 0.491 & 0.000 & 0        & N/A & N/A & 0.204    \\ % r02t_Mbh10000_Mstar1_v0_eqmass0_N1e5/ 
89           & 10000 & 0.32 & U & 0.505 & 0.465 & 0.030 & 3.45     & N/A & 842      & 0.197    \\ % r03t_Mbh10000_Mstar1_v0_eqmass0_N1e5/ 
90           & 10000 & 0.43 & U & 0.394 & 0.371 & 0.235 & 6.88     & N/A & 352      & 0.17     \\ % r04t_Mbh10000_Mstar1_v0_eqmass0_N1e5/ 
91           & 10000 & 0.50 & U & 0.288 & 0.270 & 0.442 & 7.99     & N/A & 232      & 0.122    \\ % r05t_Mbh10000_Mstar1_v0_eqmass0_N1e5/ 
92           & 10000 & 0.60 & U & 0.164 & 0.151 & 0.686 & 4.44     & N/A & 87.4     & 0.0675   \\ % r06t_Mbh10000_Mstar1_v0_eqmass0_N1e5/ 
93$^\dagger$ & 10000 & 0.65 & U & 0.127 & 0.117 & 0.756 & 5.91     & N/A & 68.7     & 0.0501   \\ % jump/r0.65t_Mbh10000_Mstar1_v0_eqmass0_N1e5_noAV_rjump/ 
94$^\dagger$ & 10000 & 0.70 & B & 0.093 & 0.085 & 0.822 & 5.60     & 8.01e+05 &0.0286   & 0.0286   \\ % jump/r0.7t_Mbh10000_Mstar1_v0_eqmass0_N1e5_noAV_rjump/ 
95$^\dagger$ & 10000 & 0.85 & B & 0.030 & 0.026 & 0.945 & 1.64     & 1.82e+05 &0.011    & 0.011    \\ % jump/r0.85t_Mbh10000_Mstar1_v0_eqmass0_N1e5_noAV_rjump/ 
96           & 10000 & 1.00 & B & 0.009 & 0.008 & 0.983 & 0.736    & 4.95e+05 &0.00245  & 0.00245  \\ % rt_Mbh10000_Mstar1_v0_eqmass0_N1e5/ 
97$^\dagger$ & 10000 & 1.00 & B & 0.011 & 0.009 & 0.980 & 0.755    & 4.91e+05 &0.00282  & 0.00282  \\ % jump/rt_Mbh10000_Mstar1_v0_eqmass0_N1e5_noAV_rjump/
\enddata
\tablecomments{\footnotesize List of all SPH calculations performed in this work. We list the initial conditions in columns 1-2 and outcomes after the first pericenter passage in the remaining columns. In column 3, we list the outcomes for a fully disrupted (F), unbound (U), and bound (B) star. In columns 4–7, we list the total mass bound to the black hole, the mass of the star, the total mass of material that has been unbound entirely from the system, and the radius of the star that encloses $99\%$ of the mass. The U and B outcomes are reported after the first pericenter passage when the separation between the star and black hole exceeds $1.7~(M_{\rm BH}/M_{\star})^{1/3} {\rm max}(r_{\rm p},r_{\rm T})$. We report the F cases at a fallback time $t_{\rm fb}$ after the first pericenter passage. In column 8, we list the orbital period of the bound, partially disrupted star to return the pericenter (in the cases of F and U outcomes, it is N/A). Finally, in columns 9-10, we list the kick velocities the star and the black hole receive: in bound cases, these values are equal and represent the kick given to the center of mass of the binary}. A $\dagger$ represents a simulation done with orbital jumps and no artificial viscosity.
\end{deluxetable*}

\startlongtable
\begin{deluxetable*}{l|rcc|ccccc|ccc}
\tabletypesize{\footnotesize}
\tablewidth{0pt}
\tablecaption{Results after last pericenter passage}
\label{last_passage}
\tablehead{
	\colhead{} &
	\colhead{$^1M_{\rm{BH}}$} &
	\colhead{$^2r_{\rm p}/r_{\rm T}$} &
	\colhead{$^3{\rm out}-$} &
	\colhead{$^4n_{\rm p}$} &
	\colhead{$^5M_{\rm{disk}}$} &
	\colhead{$^6M_{\rm{ej}}$} &
	\colhead{$^7M_{\star,f}$} &
	\colhead{$^8R_{\star,99}$} &
	\colhead{$^{9}v_{\rm{kick,rem}}$} &
	\colhead{$^{10}v_{\rm{kick,BH}}$} \\
	\colhead{} &
	\colhead{$M_{\odot}$} &
	\colhead{} &
	\colhead{\rm{come}} &
	\colhead{} &
	\colhead{$M_{\odot}$} &
	\colhead{$M_{\odot}$} &
	\colhead{$M_{\odot}$} &
	\colhead{$R_{\odot}$} &
	\colhead{km s$^{-1}$} &
	\colhead{km s$^{-1}$}
}
\startdata
1$^\dagger$ & 5     & 1.00 & F & 16 & 0.096 & 0.031 & 0.000 & 0        & N/A & 29.6     \\ % rt_Mbh5_Mstar1_v0_eqmass0_N1e5_noAV_rjump 
\hline
2           & 10    & 0.20 & F &  1 & 0.672 & 0.329 & 0.000 & 0        & N/A & 18.1     \\ % r02t_Mbh10_Mstar1_v0_eqmass0_N1e5/ 
3$^\dagger$ & 10    & 0.30 & F &  2 & 0.266 & 0.056 & 0.000 & 0        & N/A & 24.6     \\ % jump/r03t_Mbh10_Mstar1_v0_eqmass0_N1e5_noAV_rjump/ 
4$^\dagger$ & 10    & 0.43 & F &  2 & 0.294 & 0.195 & 0.000 & 0        & N/A & 27.2     \\ % jump/r04t_Mbh10_Mstar1_v0_eqmass0_N1e5_noAV_rjump/ 
5$^\dagger$ & 10    & 0.50 & U &  2 & 0.339 & 0.167 & 0.101 & 2.89     & 418      & 25.4     \\ % jump/r05t_Mbh10_Mstar1_v0_eqmass0_N1e5_noAV_rjump/ 
6$^\dagger$ & 10    & 0.60 & U &  2 & 0.354 & 0.182 & 0.222 & 2.64     & 259      & 23.9     \\ % jump/r06t_Mbh10_Mstar1_v0_eqmass0_N1e5_noAV_rjump/ 
7$^\dagger$ & 10    & 0.70 & U &  3 & 0.271 & 0.169 & 0.136 & 2.25     & 202      & 24.1     \\ % jump/r07t_Mbh10_Mstar1_v0_eqmass0_N1e5_noAV_rjump/ 
8$^\dagger$ & 10    & 0.85 & U &  6 & 0.181 & 0.116 & 0.126 & 2.15     & 124      & 23.8     \\ % jump/r085t_Mbh10_Mstar1_v0_eqmass0_N1e5_noAV_rjump/ 
9$^\dagger$ & 10    & 1.00 & U & 16 & 0.108 & 0.074 & 0.036 & 2.03     & 329      & 24.2     \\ % jump/rt_Mbh10_Mstar1_v0_eqmass0_N1e5_noAV_rjump/ 
\hline
12           & 20    & 0.20 & F &  1 & 0.554 & 0.446 & 0.000 & 0        & N/A & 15.2     \\ % jump/r02t_Mbh20_Mstar1_v0_eqmass0_N1e5/ 
13$^\dagger$ & 20    & 0.30 & F &  2 & 0.103 & 0.115 & 0.000 & 0        & N/A & 19.2     \\ % jump/r03t_Mbh20_Mstar1_v0_eqmass0_N1e5_noAV_rjump/ 
14$^\dagger$ & 20    & 0.40 & F &  2 & 0.170 & 0.187 & 0.000 & 0        & N/A & 19.7     \\ % jump/r04t_Mbh20_Mstar1_v0_eqmass0_N1e5_noAV_rjump/ 
15$^\dagger$ & 20    & 0.50 & U &  2 & 0.281 & 0.210 & 0.055 & 3.46     & 535      & 18.6     \\ % jump/r05t_Mbh20_Mstar1_v0_eqmass0_N1e5_noAV_rjump/ 
16$^\dagger$ & 20    & 0.60 & U &  2 & 0.351 & 0.210 & 0.166 & 3.24     & 371      & 17.2     \\ % jump/r06t_Mbh20_Mstar1_v0_eqmass0_N1e5_noAV_rjump/ 
17$^\dagger$ & 20    & 0.70 & U &  3 & 0.269 & 0.187 & 0.079 & 2.33     & 302      & 17.3     \\ % jump/r07t_Mbh20_Mstar1_v0_eqmass0_N1e5_noAV_rjump/ 
18$^\dagger$ & 20    & 0.85 & U &  5 & 0.132 & 0.077 & 0.402 & 3.71     & 101      & 11.6     \\ % jump/r085t_Mbh20_Mstar1_v0_eqmass0_N1e5_noAV_rjump/ 
19$^\dagger$ & 20    & 1.00 & U & 13 & 0.056 & 0.031 & 0.467 & 2.39     & 86.9     & 9.92     \\ % jump/rt_Mbh20_Mstar1_v0_eqmass0_N1e5_noAV_rjump/ 
\hline
20           & 30    & 0.20 & F &  1 & 0.505 & 0.495 & 0.000 & 0        & N/A & 13       \\ % r02t_Mbh30_Mstar1_v0_eqmass0_N1e5/ 
21           & 30    & 0.30 & U &  1 & 0.479 & 0.400 & 0.121 & 3.13     & 276      & 12.9     \\ % jump/r03t_Mbh30_Mstar1_v0_eqmass0_N1e5/ 
22           & 30    & 0.40 & U &  1 & 0.388 & 0.315 & 0.297 & 4.34     & 130      & 11       \\ % jump/r04t_Mbh30_Mstar1_v0_eqmass0_N1e5/ 
23$^\dagger$ & 30    & 0.50 & F &  2 & 0.260 & 0.258 & 0.000 & 0        & N/A & 14.9     \\ % jump/r05t_Mbh30_Mstar1_v0_eqmass0_N1e5_noAV_rjump/ 
24$^\dagger$ & 30    & 0.60 & U &  2 & 0.361 & 0.262 & 0.091 & 2.81     & 351      & 13.6     \\ % jump/r06t_Mbh30_Mstar1_v0_eqmass0_N1e5_noAV_rjump/ 
25$^\dagger$ & 30    & 0.70 & U &  3 & 0.263 & 0.190 & 0.068 & 2.57     & 356      & 13.6     \\ % jump/r07t_Mbh30_Mstar1_v0_eqmass0_N1e5_noAV_rjump/ 
26$^\dagger$ & 30    & 0.85 & U &  5 & 0.139 & 0.093 & 0.353 & 4.33     & 98.8     & 9.62     \\ % jump/r085t_Mbh30_Mstar1_v0_eqmass0_N1e5_noAV_rjump/ 
27$^\dagger$ & 30    & 1.00 & U &  9 & 0.029 & 0.015 & 0.705 & 0.841    & 29.8     & 4        \\ % jump/rt_Mbh30_Mstar1_v0_eqmass0_N1e5_noAV_rjump/ 
\hline
28           & 50    & 0.20 & F &  1 & 0.457 & 0.543 & 0.000 & 0        & N/A & 10       \\ % r02t_Mbh50_Mstar1_v0_eqmass0_N1e5/ 
29           & 50    & 0.30 & U &  1 & 0.478 & 0.442 & 0.080 & 3.12     & 569      & 9.74     \\ % r03t_Mbh50_Mstar1_v0_eqmass0_N1e5/ 
30           & 50    & 0.40 & U &  1 & 0.403 & 0.345 & 0.252 & 4.92     & 243      & 8.61     \\ % r04t_Mbh50_Mstar1_v0_eqmass0_N1e5/ 
31           & 50    & 0.50 & U &  1 & 0.282 & 0.224 & 0.494 & 5.86     & 106      & 5.94     \\ % r05t_Mbh50_Mstar1_v0_eqmass0_N1e5/ 
32$^\dagger$ & 50    & 0.60 & U &  2 & 0.354 & 0.249 & 0.098 & 3.74     & 519      & 9.6      \\ % jump/r06t_Mbh50_Mstar1_v0_eqmass0_N1e5_noAV_rjump/ 
33$^\dagger$ & 50    & 0.70 & U &  2 & 0.194 & 0.137 & 0.512 & 5.24     & 141      & 5.95     \\ % jump/r07t_Mbh50_Mstar1_v0_eqmass0_N1e5_noAV_rjump/ 
34$^\dagger$ & 50    & 0.85 & U &  4 & 0.096 & 0.061 & 0.584 & 3.01     & 134      & 5.13     \\ % jump/r085t_Mbh50_Mstar1_v0_eqmass0_N1e5_noAV_rjump/ 
35$^\dagger$ & 50    & 1.00 & U &  7 & 0.023 & 0.013 & 0.790 & 0.757    & 38       & 2.25     \\ % jump/rt_Mbh50_Mstar1_v0_eqmass0_N1e5_noAV_rjump/ 
\hline
38           & 100   & 0.22 & F &  1 & 0.423 & 0.577 & 0.000 & 0        & N/A & 6.22     \\ % r1_Mbh100_Mstar1_v0_eqmass0_N1e5/ 
39           & 100   & 0.32 & U &  1 & 0.457 & 0.488 & 0.054 & 3.38     & 705      & 6.09     \\ % r1.6_Mbh100_Mstar1_v0_eqmass0_N1e5/ 
40           & 100   & 0.43 & U &  1 & 0.373 & 0.337 & 0.289 & 6.41     & 271      & 5.08     \\ % r2_Mbh100_Mstar1_v0_eqmass0_N1e5/ 
41           & 100   & 0.54 & U &  1 & 0.234 & 0.199 & 0.567 & 6.13     & 131      & 3.31     \\ % r2.5_Mbh100_Mstar1_v0_eqmass0_N1e5/ 
42           & 100   & 0.60 & U &  1 & 0.170 & 0.135 & 0.696 & 4.67     & 43.6     & 2.01     \\ % r2.9_Mbh100_Mstar1_v0_eqmass0_N1e5/ 
43$^\dagger$ & 100   & 0.65 & U &  2 & 0.273 & 0.217 & 0.285 & 6.10     & 273      & 4.99     \\ % jump/r0.65t_Mbh100_Mstar1_v0_eqmass0_N1e5_noAV_rjump/ 
44$^\dagger$ & 100   & 0.70 & U &  2 & 0.191 & 0.146 & 0.502 & 5.81     & 172      & 3.71     \\ % jump/r0.7t_Mbh100_Mstar1_v0_eqmass0_N1e5_noAV_rjump/ 
45$^\dagger$ & 100   & 0.85 & U &  3 & 0.070 & 0.049 & 0.733 & 1.67     & 27.3     & 1.66     \\ % jump/r0.85t_Mbh100_Mstar1_v0_eqmass0_N1e5_noAV_rjump/ 
47$^\dagger$ & 100   & 1.00 & U &  5 & 0.021 & 0.012 & 0.859 & 0.742    & 36.9     & 0.992    \\ % jump/rt_Mbh100_Mstar1_v0_eqmass0_N1e5_noAV_rjump/ 
48$^\dagger$ & 100   & 1.10 & U & 10 & 0.009 & 0.005 & 0.868 & 0.641    & 13.5     & 0.759    \\ % jump/r1.10t_Mbh100_Mstar1_v0_eqmass0_N1e5_noAV_rjump/ 
\hline
52           & 200   & 0.17 & F &  1 & 0.384 & 0.616 & 0.000 & 0        & N/A & 3.71     \\ % r1_Mbh200_Mstar1_v0_eqmass0_N1e5_bettercns/ 
53           & 200   & 0.32 & U &  1 & 0.438 & 0.522 & 0.040 & 3.69     & 533      & 3.57     \\ % r1.9_Mbh200_Mstar1_v0_eqmass0_N1e5/ 
54           & 200   & 0.43 & U &  1 & 0.373 & 0.372 & 0.256 & 7.72     & 316      & 3.07     \\ % r2.5_Mbh200_Mstar1_v0_eqmass0_N1e5/ 
55           & 200   & 0.51 & U &  1 & 0.260 & 0.247 & 0.493 & 7.05     & 186      & 2.25     \\ % r3_Mbh200_Mstar1_v0_eqmass0_N1e5/ 
56           & 200   & 0.60 & U &  1 & 0.162 & 0.147 & 0.691 & 4.27     & 80.4     & 1.35     \\ % r3.5_Mbh200_Mstar1_v0_eqmass0_N1e5/ 
57$^\dagger$ & 200   & 0.65 & U &  1 & 0.139 & 0.098 & 0.763 & 5.94     & 58.2     & 0.984    \\ % jump/r0.65t_Mbh200_Mstar1_v0_eqmass0_N1e5_noAV_rjump/ 
58$^\dagger$ & 200   & 0.70 & U &  2 & 0.188 & 0.153 & 0.496 & 6.84     & 160      & 2.14     \\ % jump/r0.7t_Mbh200_Mstar1_v0_eqmass0_N1e5_noAV_rjump/ 
59$^\dagger$ & 200   & 0.85 & U &  3 & 0.068 & 0.050 & 0.734 & 1.95     & 111      & 1.25     \\ % jump/r0.85t_Mbh200_Mstar1_v0_eqmass0_N1e5_noAV_rjump/ 
61$^\dagger$ & 200   & 1.00 & U &  6 & 0.021 & 0.015 & 0.824 & 0.755    & 55.6     & 0.771    \\ % jump/rt_Mbh200_Mstar1_v0_eqmass0_N1e5_noAV_rjump/ 
62$^\dagger$ & 200   & 1.10 & U &  8 & 0.009 & 0.005 & 0.895 & 0.647    & 44.4     & 0.486    \\ % jump/r1.10t_Mbh200_Mstar1_v0_eqmass0_N1e5_noAV_rjump/ 
\hline
66           & 500   & 0.25 & F &  1 & 0.362 & 0.638 & 0.000 & 0        & N/A & 1.79     \\ % r2_Mbh500_Mstar1_v0_eqmass0_N1e5/ 
67           & 500   & 0.32 & U &  1 & 0.506 & 0.456 & 0.038 & 3.12     & 536      & 1.86     \\ % r2.5_Mbh500_Mstar1_v0_eqmass0_N1e5/ 
68           & 500   & 0.38 & U &  1 & 0.414 & 0.511 & 0.074 & 6.93     & 444      & 1.66     \\ % r3_Mbh500_Mstar1_v0_eqmass0_N1e5/ 
69           & 500   & 0.50 & U &  1 & 0.268 & 0.271 & 0.460 & 7.58     & 216      & 1.16     \\ % r4_Mbh500_Mstar1_v0_eqmass0_N1e5/ 
70           & 500   & 0.63 & U &  1 & 0.140 & 0.118 & 0.742 & 4.11     & 61.3     & 0.527    \\ % r5_Mbh500_Mstar1_v0_eqmass0_N1e5/ 
71$^\dagger$ & 500   & 0.65 & U &  1 & 0.136 & 0.107 & 0.758 & 6.57     & 69.3     & 0.499    \\ % jump/r0.65t_Mbh500_Mstar1_v0_eqmass0_N1e5_noAV_rjump/ 
72$^\dagger$ & 500   & 0.70 & U &  2 & 0.182 & 0.155 & 0.496 & 6.72     & 192      & 1.05     \\ % jump/r0.7t_Mbh500_Mstar1_v0_eqmass0_N1e5_noAV_rjump/ 
73$^\dagger$ & 500   & 0.85 & U &  4 & 0.092 & 0.075 & 0.569 & 3.18     & 135      & 0.883    \\ % jump/r0.85t_Mbh500_Mstar1_v0_eqmass0_N1e5_noAV_rjump/ 
75$^\dagger$ & 500   & 1.00 & U &  6 & 0.021 & 0.015 & 0.825 & 0.776    & 35.4     & 0.339    \\ % jump/rt_Mbh500_Mstar1_v0_eqmass0_N1e5_noAV_rjump/ 
76$^\dagger$ & 500   & 1.10 & U &  6 & 0.008 & 0.005 & 0.924 & 0.657    & 29.1     & 0.162    \\ % jump/r1.10t_Mbh500_Mstar1_v0_eqmass0_N1e5_noAV_rjump/ 
\hline
77           & 1000  & 0.20 & F &  1 & 0.469 & 0.531 & 0.000 & 0        & N/A & 1.09     \\ % r2_Mbh1000_Mstar1_v0_eqmass0_N1e5_c1_0.02/ 
78           & 1000  & 0.30 & U &  1 & 0.445 & 0.540 & 0.015 & 4.87     & 915      & 0.968    \\ % r3_Mbh1000_Mstar1_v0_eqmass0_N1e5/ 
79           & 1000  & 0.40 & U &  1 & 0.346 & 0.514 & 0.140 & 11.5     & 396      & 0.963    \\ % r4_Mbh1000_Mstar1_v0_eqmass0_N1e5/ 
80           & 1000  & 0.50 & U &  1 & 0.280 & 0.274 & 0.445 & 7.35     & 226      & 0.681    \\ % r5_Mbh1000_Mstar1_v0_eqmass0_N1e5/ 
81           & 1000  & 0.60 & U &  1 & 0.159 & 0.155 & 0.686 & 5.45     & 101      & 0.38     \\ % r6_Mbh1000_Mstar1_v0_eqmass0_N1e5/ 
82$^\dagger$ & 1000  & 0.65 & U &  1 & 0.134 & 0.111 & 0.755 & 7.38     & 64.1     & 0.282    \\ % jump/r0.65t_Mbh1000_Mstar1_v0_eqmass0_N1e5_noAV_rjump/ 
83$^\dagger$ & 1000  & 0.70 & U &  2 & 0.176 & 0.156 & 0.498 & 7.73     & 181      & 0.59     \\ % jump/r0.7t_Mbh1000_Mstar1_v0_eqmass0_N1e5_noAV_rjump/ 
84$^\dagger$ & 1000  & 0.85 & U &  3 & 0.063 & 0.051 & 0.739 & 1.54     & 137      & 0.348    \\ % jump/r0.85t_Mbh1000_Mstar1_v0_eqmass0_N1e5_noAV_rjump/ 
86$^\dagger$ & 1000  & 1.00 & U &  6 & 0.020 & 0.016 & 0.826 & 0.736    & 33.8     & 0.193    \\ % jump/rt_Mbh1000_Mstar1_v0_eqmass0_N1e5_noAV_rjump/ 
87$^\dagger$ & 1000  & 1.10 & U &  9 & 0.008 & 0.006 & 0.884 & 0.647    & 38.1     & 0.137    \\ % jump/r1.10t_Mbh1000_Mstar1_v0_eqmass0_N1e5_noAV_rjump/ 
\hline
88           & 10000 & 0.20 & F &  1 & 0.509 & 0.491 & 0.000 & 0        & N/A & 0.204    \\ % r02t_Mbh10000_Mstar1_v0_eqmass0_N1e5/ 
89           & 10000 & 0.32 & U &  1 & 0.505 & 0.465 & 0.030 & 3.45     & 842      & 0.197    \\ % r03t_Mbh10000_Mstar1_v0_eqmass0_N1e5/ 
90           & 10000 & 0.43 & U &  1 & 0.394 & 0.371 & 0.235 & 6.88     & 352      & 0.17     \\ % r04t_Mbh10000_Mstar1_v0_eqmass0_N1e5/ 
91           & 10000 & 0.50 & U &  1 & 0.288 & 0.270 & 0.442 & 7.99     & 232      & 0.122    \\ % r05t_Mbh10000_Mstar1_v0_eqmass0_N1e5/ 
92           & 10000 & 0.60 & U &  1 & 0.164 & 0.151 & 0.686 & 4.44     & 87.4     & 0.0675   \\ % r06t_Mbh10000_Mstar1_v0_eqmass0_N1e5/ 
93$^\dagger$ & 10000 & 0.65 & U &  1 & 0.127 & 0.117 & 0.756 & 5.91     & 68.7     & 0.0501   \\ % jump/r0.65t_Mbh10000_Mstar1_v0_eqmass0_N1e5_noAV_rjump/ 
94$^\dagger$ & 10000 & 0.70 & U &  2 & 0.162 & 0.159 & 0.502 & 11.8     & 176      & 0.0873   \\ % jump/r0.7t_Mbh10000_Mstar1_v0_eqmass0_N1e5_noAV_rjump/ 
97$^\dagger$ & 10000 & 1.00 & U &  3 & 0.015 & 0.014 & 0.925 & 0.714    & 20.9     & 0.0127   \\ % jump/rt_Mbh10000_Mstar1_v0_eqmass0_N1e5_noAV_rjump/
\enddata
\tablecomments{\footnotesize 
List of all SPH calculations performed in this work. We list the initial conditions in columns 1-2 and outcomes after the last pericenter passage in the remaining columns. In columns 3-4, we list the final outcomes and the total number of passages. In columns 5–8, we list the total mass bound to the black hole, the mass of the star, the mass of material that has been unbound entirely from the system, and the radius of the star that encloses $99\%$ of the mass, with all these outcomes are reported at $1t_{\rm fb}$ after the last pericenter passage. Columns 9-10 report the final kick velocities given to the remnant and the BH, respectively.
 A $\dagger$ represents a simulation done with orbital jumps and no artificial viscosity: during the orbital jumps of these calculations, the mass bound to the black hole is added to the black hole mass and the mass unbound from the system is excised from the calculation. In these cases, the disk mass $M_{\rm disk}$ and ejecta mass $M_{\rm ej}$ listed in this table include only the mass stripped on the final pericenter passage, and thus the sum of columns 5, 6, and 7 is less than the initial star mass of 1 $M_\odot$.}
\end{deluxetable*}

\bibliography{refs}

\end{document}